\documentclass{jfm}
\usepackage{graphicx}
\usepackage{epstopdf, epsfig}
\usepackage{amsmath}
\usepackage{color}
\usepackage{hyperref}
\usepackage{url}

\hypersetup{
    colorlinks=true,
    linkcolor=blue,
    filecolor=blue,
    urlcolor=blue,
    citecolor=blue,
}

\shorttitle{Skin friction and heat transfer}
\shortauthor{D. Xu, J. Wang and S. Chen}

\title{Skin friction and heat transfer in hypersonic transitional and turbulent boundary layers}

\author{Dehao Xu\aff{1},
	Jianchun Wang\aff{2}
\corresp{\email{wangjc@sustech.edu.cn}}
\and Shiyi Chen \aff{1,2}
\corresp{\email{chensy@sustech.edu.cn}}}
		
\affiliation{\aff{1}State Key Laboratory of Turbulence and Complex Systems, College of Engineering, Peking University, Beijing 100871, People$^{'}$s Republic of China
\aff{2}Department of Mechanics and Aerospace Engineering, Southern University of Science and Technology, Shenzhen 518055, People$^{'}$s Republic of China}

\begin{document}

\maketitle

\begin{abstract}
The decompositions of the skin-friction and heat transfer coefficients based on the two-fold repeated integration in hypersonic transitional and turbulent boundary layers are analyzed to explain the generations of the wall skin friction and heat transfer. The Reynolds analogy factor slightly increases as the wall temperature decreases, especially for the extremely cooled wall. The integral analysis is applied to explain the overshoot behaviours of the skin-friction and heat transfer coefficients in hypersonic transitional boundary layers. The overshoot of the skin-friction coefficient is mainly caused by the drastic change of the mean velocity profiles, and the overshoot of the heat transfer coefficient is primarily due to the viscous dissipation. In the hypersonic turbulent boundary layers, the skin-friction and heat transfer coefficients increase significantly as the wall temperature decreases. The effects of the mean velocity gradients and the Reynolds shear stress contribute dominantly to the wall skin friction, and have weak correlations with the wall temperature, except for the strongly cooled wall condition. The strongly cooled wall condition and high Mach number can enhance the effect of the Reynolds shear stress, and weaken the impact of the mean velocity gradients. Furthermore, the magnitudes of the dominant relative contributions of the mean temperature gradients, pressure dilatation, viscous dissipation and the Reynolds heat flux to the heat transfer coefficient increase as the wall temperature increases in the hypersonic turbulent boundary layers.
\end{abstract}


\section{Introduction}
Understanding the underlying mechanism of the mean skin friction generation in wall-bounded turbulence is of great importance, particularly for the design of the skin friction reduction in engineering applications. Furthermore, the heat transfer is also an important topic in compressible wall-bounded turbulence, and the potential mechanism of the heat transfer generation is significant for the thermal protection in the supersonic and hypersonic aircrafts.

In the past decades, plenty of works were devoted to express the skin friction as the contributions from the flow statistics inside the flow domain in wall-bounded turbulence \citep[]{Fukagata2002,Gomez2009,Renard2016,Li2019}. \citet[]{Fukagata2002} first derived a simple relationship between the skin-friction coefficient and the componential contributions from the mean and statistical turbulent quantities across the wall layer, and this relationship was referred to the FIK identity. The FIK identity decomposes the skin-friction coefficient into the laminar, turbulent, inhomogeneous and transient components based on the streamwise momentum budget. Furthermore, several modifications to the FIK identity have been proposed in previous investigations. \citet[]{Peet2009} and \citet[]{Bannier2015} generalized the FIK identity to complex geometries to investigate the skin friction reduction with riblets. \citet[]{Modesti2018} derived a generalized form of the FIK identity in order to quantify the effect of cross-stream convection on the mean streamwise velocity, wall shear stress and bulk friction coefficient in turbulent square duct flows. Furthermore, \citet[]{Mehdi2011} modified the FIK identity by replacing the streamwise gradients with total stress gradients in the wall-normal direction, in order to avoid the significant uncertainties of the experimental measurements of the streamwise gradients. \citet[]{Mehdi2014} proposed another modified form of the FIK identity in which the upper integration limit can be the arbitrary wall-normal location with the boundary layer, in order to deal with the flows with ill-defined outer boundary conditions or when the measurement grid does not extend over the whole boundary layer. Moreover, the FIK identity was extended by  \citet[]{Gomez2009} to compressible wall-bounded turbulence and was used to studied the compressibility effects on the generation of skin friction. They showed that the skin friction can be ascribed to the contributions of four physical processes, i.e. the laminar, turbulent, compressible and a fourth coming from the interaction between turbulence and compressibility. They found that the main contribution to the skin friction is from the turbulent term in compressible turbulent channel flows.

A drawback of the FIK identity is that some of the contributing terms can not be easily interpreted for the three successive integrations. Particularly, there is no physics-informed explanation for the linearly weighted Reynolds shear stress. Therefore, \citet[]{Renard2016} proposed another decomposition of skin friction based on the mean streamwise kinetic-energy equation in an absolute reference frame, in which the undisturbed fluid is not moving, and this new decomposition was named as the RD identity. They pointed out that the RD identity gives an improved physical interpretation of the potential mechanism of the generation of the skin friction. Recently, the RD identity was extended to compressible turbulent channel flows \citep[]{Li2019} and zero-pressure-gradient boundary flows \citep[]{Fan2019}.

However, when it comes to the decomposition of the heat transfer coefficient in compressible wall-bounded turbulence, a similar derivation procedure of the RD identity in the skin-friction coefficient is hard to be applied on the temperature or total-enthalpy equation. From this perspective, an integral method similar to the FIK identity is more applicable compared to the RD identity. According to the viewpoints of \citet[]{Zhang2020} and \citet[]{Wenzel2021b}, the two-fold repeated integration is better than the three-fold repeated integration in the FIK identity. They showed that the decomposition based on the two-fold repeated integration gets rid of the linearly weighted term on the Reynolds shear stress and has more intuitive physical interpretation of the underlying mechanisms of the generation of the skin-friction and heat transfer coefficients. \citet[]{Zhang2020} used this new decomposition to investigate the heat transfer coefficient in supersonic turbulent channel flows. They found that the viscous dissipation gives the dominant contribution to the heat transfer coefficient. \citet[]{Wenzel2021b} applied this decomposition to investigate the skin-friction and heat transfer coefficients in subsonic and supersonic turbulent boundary layers. They investigated the effect of Mach number, wall temperature and pressure gradient on the decomposed components of the skin-friction and heat transfer coefficients. However, to the best of our knowledge, the integral analysis was all applied on the decomposition of the skin-friction and heat transfer coefficients in turbulent wall-bounded flows, and was not employed to study the transitional wall-bounded flows, in which the underlying mechanisms of the skin friction and wall heat transfer are unclear and of fundamental and practical importance in the aerospace industry. Furthermore, it should be noted that one of the most important differences between supersonic and hypersonic turbulent boundary layers is the wall temperature condition \citep{Duan2010}. If an aircraft flies at a supersonic speed, the wall temperature is essentially considered as adiabatic, while for an aircraft at a hypersonic speed, the wall temperature is remarkably lower than the adiabatic wall temperature \citep{Duan2010}. It was found that the cooled wall temperature can result in strong wall heat transfer \citep{Zhang2014,Xu2021b}. Accordingly, it is urgent to investigate the potential mechanisms of the skin friction and wall heat transfer and the effect of wall temperature in hypersonic turbulent boundary layers.

The physical mechanism of laminar to turbulent transition for compressible wall-bounded flows has not been well understood. It has been widely found that the skin friction and heat transfer values first overshoot the turbulent values in the nonlinear breakdown regime before ultimately decrease to the turbulent boundary layer values \citep{Brandt2002,Horvath2002,Pirozzoli2004,Wadhams2008,Mayer2011,Franko2011,Franko2013,Liang2013,Liang2015}. The physical mechanisms of the overshoot phenomena in the skin-friction and heat transfer coefficients are still unclear and lead to a great challenge to the prediction of the transition to turbulence. \citet{Franko2013} tried to figure out the mechanism of the overshoot in a Mach 6 transitional boundary layer. They inferred that the disturbances lead to the generation of the strong streamwise vorticity and the streaks, and the increased transport in the wall-normal direction of thermal energy and momentum due to the streamwise vorticity causes the overshoot of the skin-friction and heat transfer coefficients. The Reynolds shear stress and Reynolds heat flux terms responsible for the transport of momentum and thermal energy lead to the increase of skin friction and wall heat transfer. However, it is still unclear which specific mechanism is responsible for the overshoot of wall heat transfer in high-speed boundary layer.

According to the above introduction, it is found that the underlying mechanisms of the skin-friction and heat transfer coefficients in transitional wall-bounded flows have never been studied using the decomposition methods. Furthermore, previous investigations of the decompositions of the skin-friction and heat transfer coefficients were limited to the subsonic and supersonic turbulent boundary layers, where the compressibility effect and heat transfer near the cooled wall are much weaker than those in hypersonic turbulent boundary layers. Accordingly, whether the mechanisms of skin friction and wall heat transfer will be different in hypersonic turbulent boundary layers are still unclear. Therefore, the main goal of this study is applying the decomposition methods to investigate the potential mechanisms of the skin-friction and heat transfer coefficients in hypersonic transitional and turbulent boundary layers. The effect of wall temperature on the decompositions of the skin-friction and heat transfer coefficients is also discussed. Furthermore, the reasons of the overshoot phenomena of the skin-friction and heat transfer coefficients are given by the integral analysis for the first time.

The rest of the paper is organized as follows. The governing equations and simulation parameters are described in Section \ref{sec: n1}. The mathematical forms of the decompositions of the skin-friction and heat transfer coefficients are derived in Section \ref{sec: n2}. The streamwise evolutions of the skin friction and wall heat transfer in hypersonic transitional and turbulent boundary layers are presented in Section \ref{sec: n3} and \ref{sec: n4} respectively. Finally, summary and conclusions are given in Section \ref{sec: n5}.

\section{Governing equations and simulation parameters}\label{sec: n1}
The compressible Navier-Stokes equations are non-dimensionalized by the following set of reference scales, including the reference length $L_{\infty }$, free-stream density $\rho  _{\infty }$, velocity $U_{\infty }$, temperature $T_{\infty}$, pressure $p_{\infty }= \rho  _{\infty }U_{\infty }^{2}$, energy per unit volume $\rho  _{\infty }U_{\infty }^{2}$, viscosity $\mu _{\infty }$ and thermal conductivity $\kappa _{\infty }$. Three non-dimensional governing parameters are introduced, namely the Reynolds number $ Re= \rho _{\infty }U_{\infty }L_{\infty }/\mu _{\infty }$, the Mach number $ M= U_{\infty }/c_{\infty }$ and Prandtl number $ Pr= \mu _{\infty } C_{p}/\kappa _{\infty }$. The ratio of specific heat at constant pressure $C_{p}$ to that at constant volume $C_{v}$ is defined as $\gamma = C_{p}/C_{v}$ and assumed to be equal to 1.4. The parameter $ \alpha$ is given by $ \alpha = PrRe\left ( \gamma -1 \right )M^{2}$, where $Pr=0.7$.

The following compressible dimensionless Navier-Stokes equations in the conservative form are solved numerically \citep[]{Liang2015,Xu2021}
\begin{equation}
		\frac{\partial \rho }{\partial t}+\frac{\partial \left ( \rho u_{j} \right )}{\partial x_{j}}=0,
\end{equation}
\begin{equation}
		\frac{\partial \left ( \rho u_{i} \right )}{\partial t}+\frac{\partial \left [ \rho u_{i}u_{j}+p\delta _{ij} \right ]}{\partial x_{j}}=\frac{1}{Re}\frac{\partial \sigma _{ij}}{\partial x_{j}},
\end{equation}
\begin{equation}
		\frac{\partial E}{\partial t}+\frac{\partial \left [ \left ( E+p \right )u_{j} \right ]}{\partial x_{j}}=\frac{1}{\alpha }\frac{\partial }{\partial x_{j}}\left ( \kappa \frac{\partial T}{\partial x_{j}} \right )+\frac{1}{Re}\frac{\partial \left ( \sigma _{ij}u_{i} \right )}{\partial x_{j}},
\end{equation}
\begin{equation}
		p=\rho T/\left ( \gamma M^{2} \right ),
\end{equation}
where $\rho $, $u_{i}$, $T$ and $p$ are the density, velocity component, temperature and pressure, respectively. The viscous stress $\sigma _{ij}$ is defined as
\begin{equation}
\sigma _{ij}=\mu \left ( \frac{\partial u_{i}}{\partial x_{j}}+\frac{\partial u_{j}}{\partial x_{i}} \right )-\frac{2}{3}\mu \theta \delta _{ij},
\end{equation}
where $\theta = \partial u_{k}/\partial x_{k}$ is the velocity divergence. The total energy per unit volume $E$ is
\begin{equation}
E=\frac{p}{\gamma -1}+\frac{1}{2}\rho \left ( u_{j}u_{j} \right ).
\end{equation}

The convection terms of the compressible governing equations are approximated by a seventh-order weighted essentially non-oscillatory scheme \citep[]{Balsara2000}, and the viscous terms are discretized by an eighth-order central difference scheme. A third-order total variation diminishing type of Runge-Kutta method is applied for time advancing \citep{shu1988,Liang2013}.

In this study, $\overline{f}$ denotes the Reynolds average (spanwise and time average) of $f$, and the fluctuating component of the Reynolds average is ${f}'=f-\overline{f}$. Moreover, $\widetilde{f}=\overline{\rho f}/\bar{\rho }$ represents the Favre average of $f$, and the fluctuating component is ${f}''=f-\widetilde{f}$.

\begin{figure}\centering
         \includegraphics[width=0.48\linewidth]{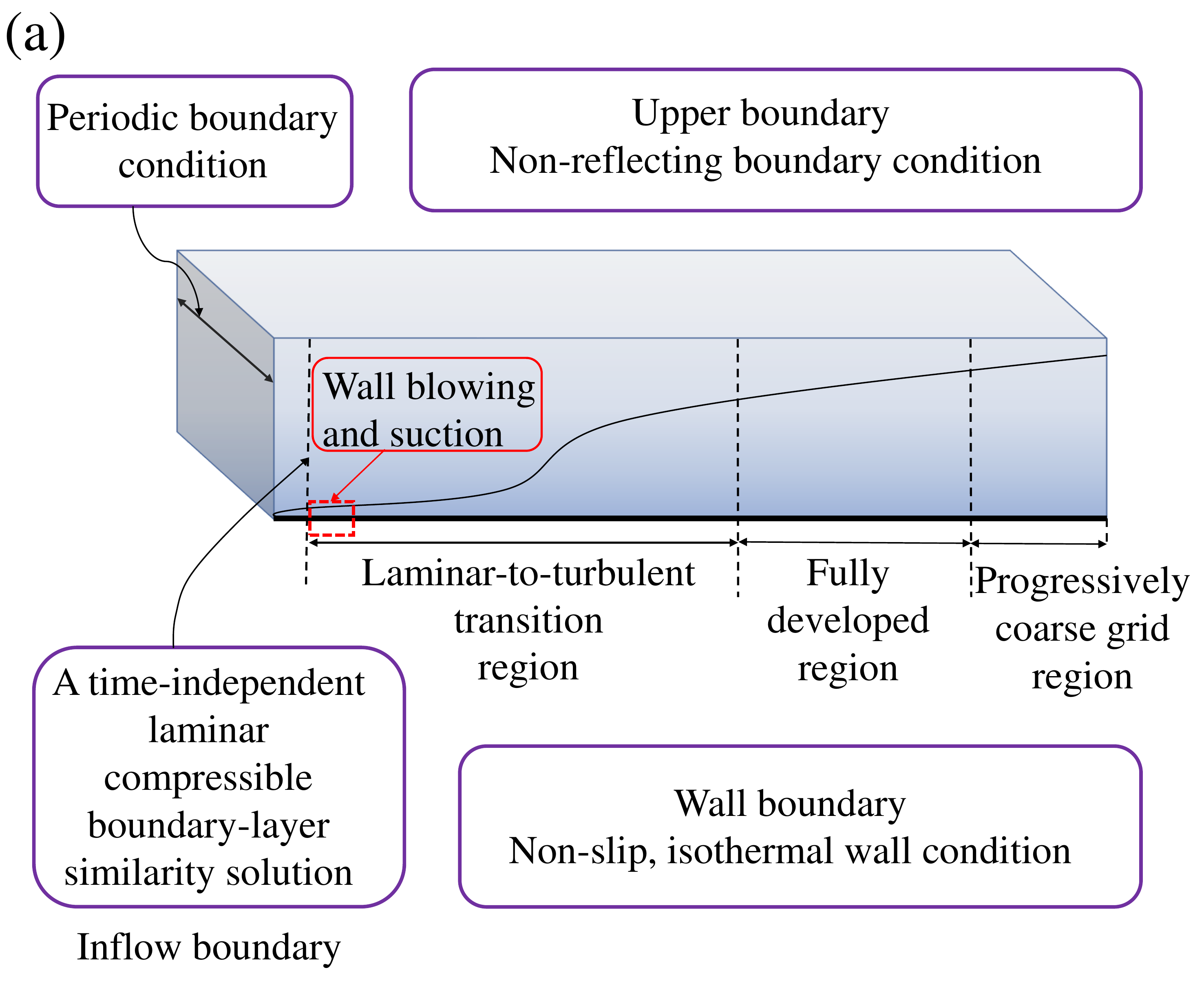}
         \includegraphics[width=0.48\linewidth]{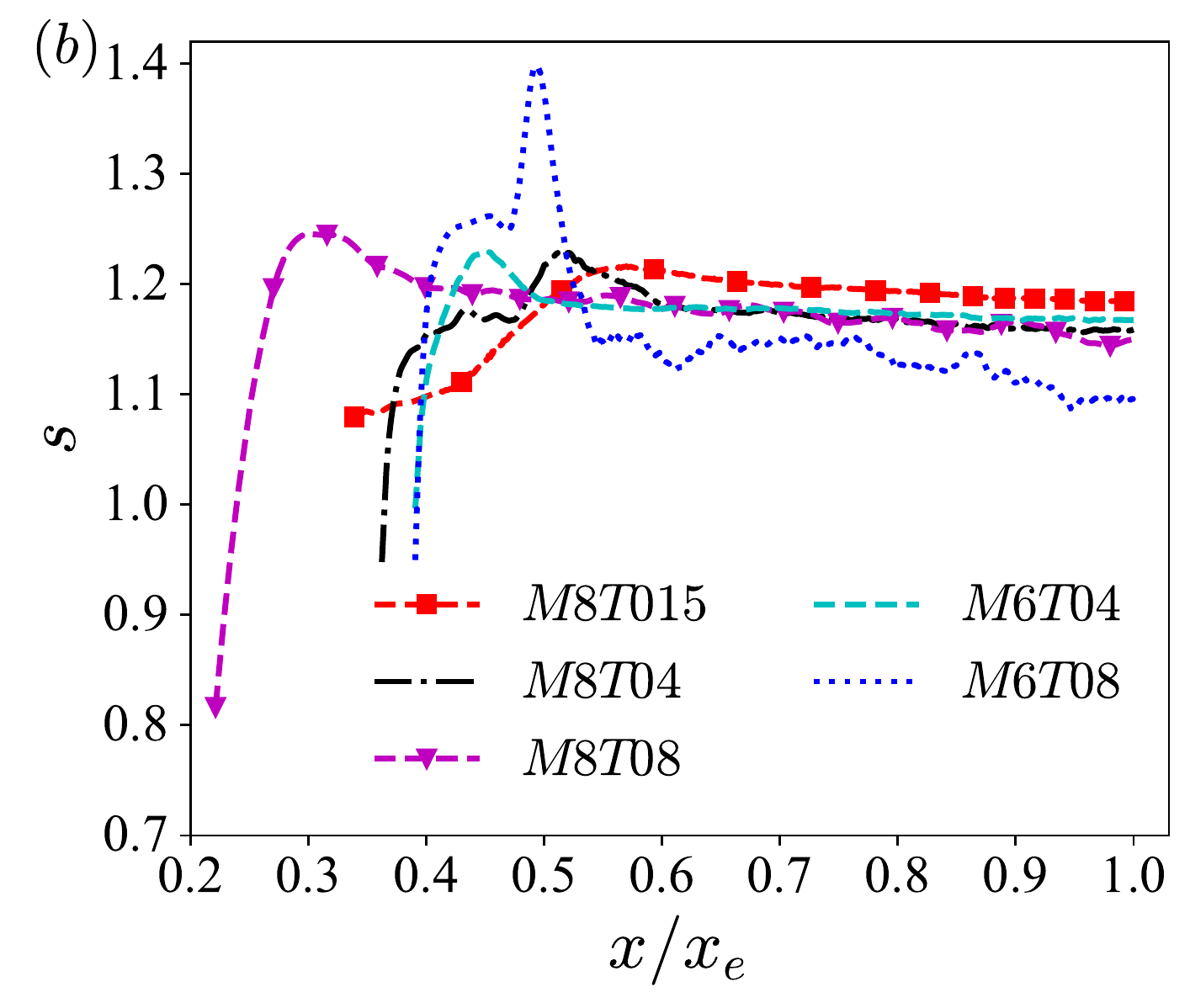}
	\caption{(a) A schematic of the hypersonic transitional and turbulent boundary layers. (b) The Reynolds analogy factor $s$ along the streamwise direction in the hypersonic transitional and turbulent boundary layers.}
	\label{fig: d1}
\end{figure}

The DNS is performed using the OPENCFD code, which has been widely validated in compressible transitional and turbulent wall-bounded flows \citep[]{Liang2013,Liang2015,Zhang2014,She2018,Xu2021,Xu2021b}. A schematic of the hypersonic transitional and turbulent boundary layers is shown in figure \ref{fig: d1} (a). The spatially evolving hypersonic transitional and turbulent boundary layers are numerically simulated with the following boundary conditions: the inflow and outflow boundary conditions, a wall boundary condition, an upper far-field boundary condition, and a periodic boundary condition in the spanwise direction. To be specific, a time-independent laminar compressible boundary-layer similarity solution is applied at the inflow boundary. A region of wall blowing and suction is implemented to induce the laminar-to-turbulent transition. The blowing and suction disturbance applied on the wall-normal velocity component on the wall can be defined as \citep[]{Rai1995,Pirozzoli2004}
\begin{equation}
	v\left ( x,z,t \right )|_{wall}=AU_{\infty }f\left ( x \right )g\left ( z \right )h\left ( t \right ),\,x_{a}\leq x\leq x_{b}.
\end{equation}
Here $A$ is the amplitude of the disturbance and $ U_{\infty }$ is the free-stream streamwise velocity. $ x_{a}=4.5$ and $ x_{b}=5.0$ are the starting and ending positions of the blowing and suction zone respectively. Furthermore,
\begin{equation}
	f\left ( x \right )=4sin\theta \left ( 1-cos\theta  \right )/\sqrt{27},\,\theta =2\pi \left ( x-x_{a} \right )/\left ( x_{b}-x_{a} \right );
\end{equation}
\begin{equation}
	g\left ( z \right )=\sum_{l=1}^{l_{max}}Z_{l}sin\left [ 2\pi l \left (z/L_{z}+\phi _{l}  \right ) \right ],\,\sum_{l=1}^{l_{max}}Z_{l}=1,\,Z_{l}=1.25Z_{l+1};
\end{equation}
\begin{equation}
	h\left ( t \right )=\sum_{m=1}^{m_{max}}T_{m}sin\left [ 2\pi m \left (\beta t+\phi _{m}  \right ) \right ],\,\sum_{m=1}^{m_{max}}T_{m}=1,\,T_{m}=1.25T_{m+1}.
\end{equation}
Here $L_{z}$ is the size of the spanwise domain and $l_{max}=10$, $ m_{max}=5$. $\phi _{l}$ and $\phi _{m}$ are the phase difference, and taken as random numbers ranging between 0 and 1. The values of $A$ and $\beta $ of database are listed in table \ref{tab: tab1}.

\begin{table}
	\begin{center}
		\def~{\hphantom{0}}
\setlength{\tabcolsep}{6mm}
		\begin{tabular}{ccc}
			Case & $A$ &$\beta $\\ [4pt]
            M6T04 & 0.1 & -1.0 \\
			M6T08 & 0.15 &-1.0 \\
			M8T015 & 0.04 &1.57 \\
			M8T04 & 0.15 & -1.0 \\
			M8T08 &0.08 &0.5 \\
		\end{tabular}
		\caption{The values of $A$ and $\beta $ of the DNS study.}
		\label{tab: tab1}
	\end{center}
\end{table}

Moreover, in order to inhibit the reflection of disturbance due to the numerical treatment of the outflow boundary condition, a progressively coarse grid is implemented in the streamwise direction near the outflow boundary condition \citep[]{Pirozzoli2004}. The non-slip and isothermal boundary conditions are applied for the wall boundary, and the non-reflecting boundary condition is imposed for the upper boundary. Accordingly, the hypersonic transitional and turbulent boundary layer initially undergoes a bypass transition in the laminar-to-turbulent transition region, and then achieves the fully developed state in the fully developed region.

\begin{figure}\centering
         \includegraphics[width=0.96\linewidth]{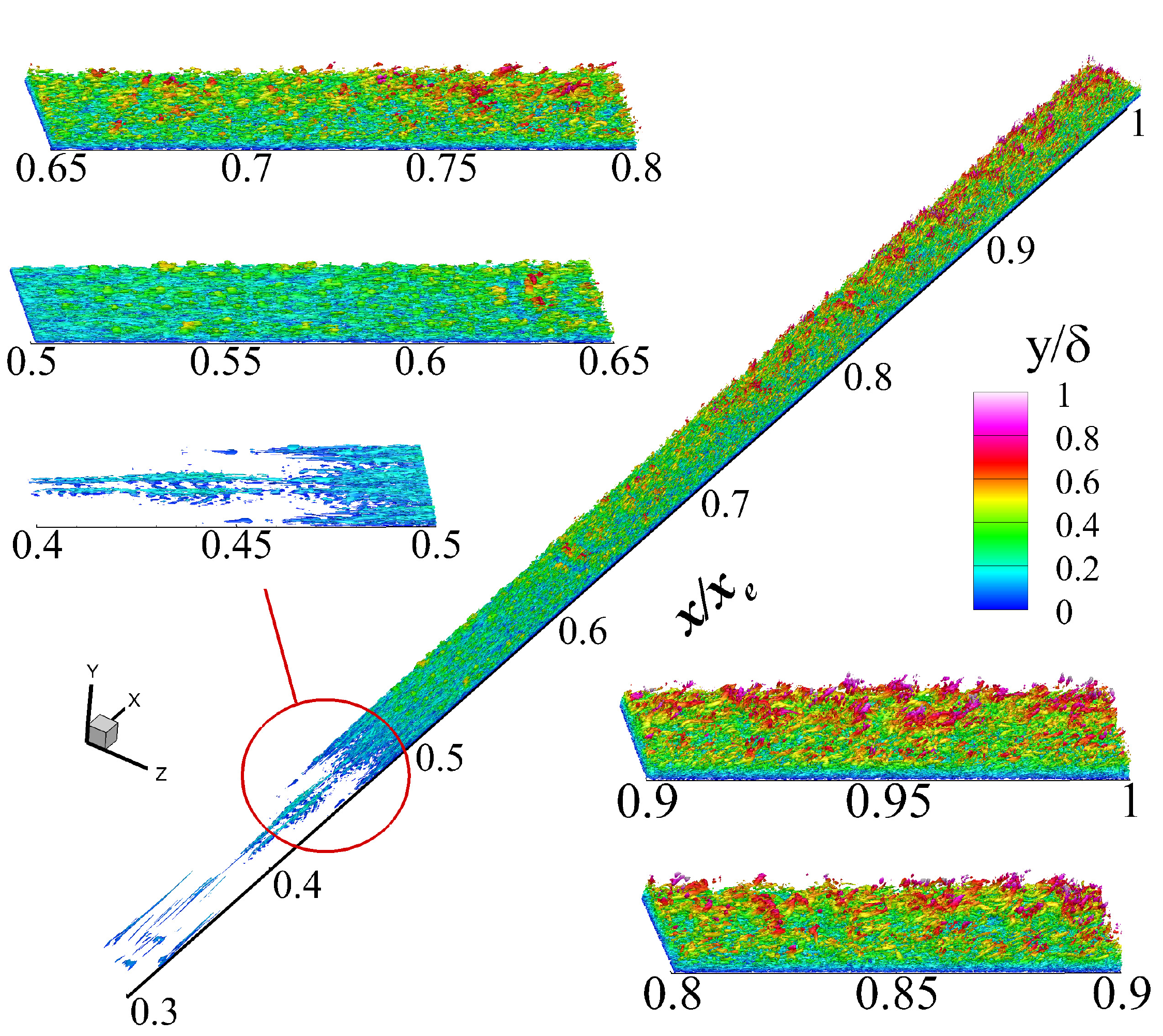}
	\caption{Visualization of the instantaneous vortical structures based on the Q-criterion \citep[]{Hunt1988} of the hypersonic transitional and turbulent boundary layer in M8T015. The structures are coloured by the wall-normal location $y/\delta$. The isosurfaces of the instantaneous vortical structures represent $Q=10$.}
	\label{fig: d2}
\end{figure}

\begin{figure}\centering
         \includegraphics[width=0.96\linewidth]{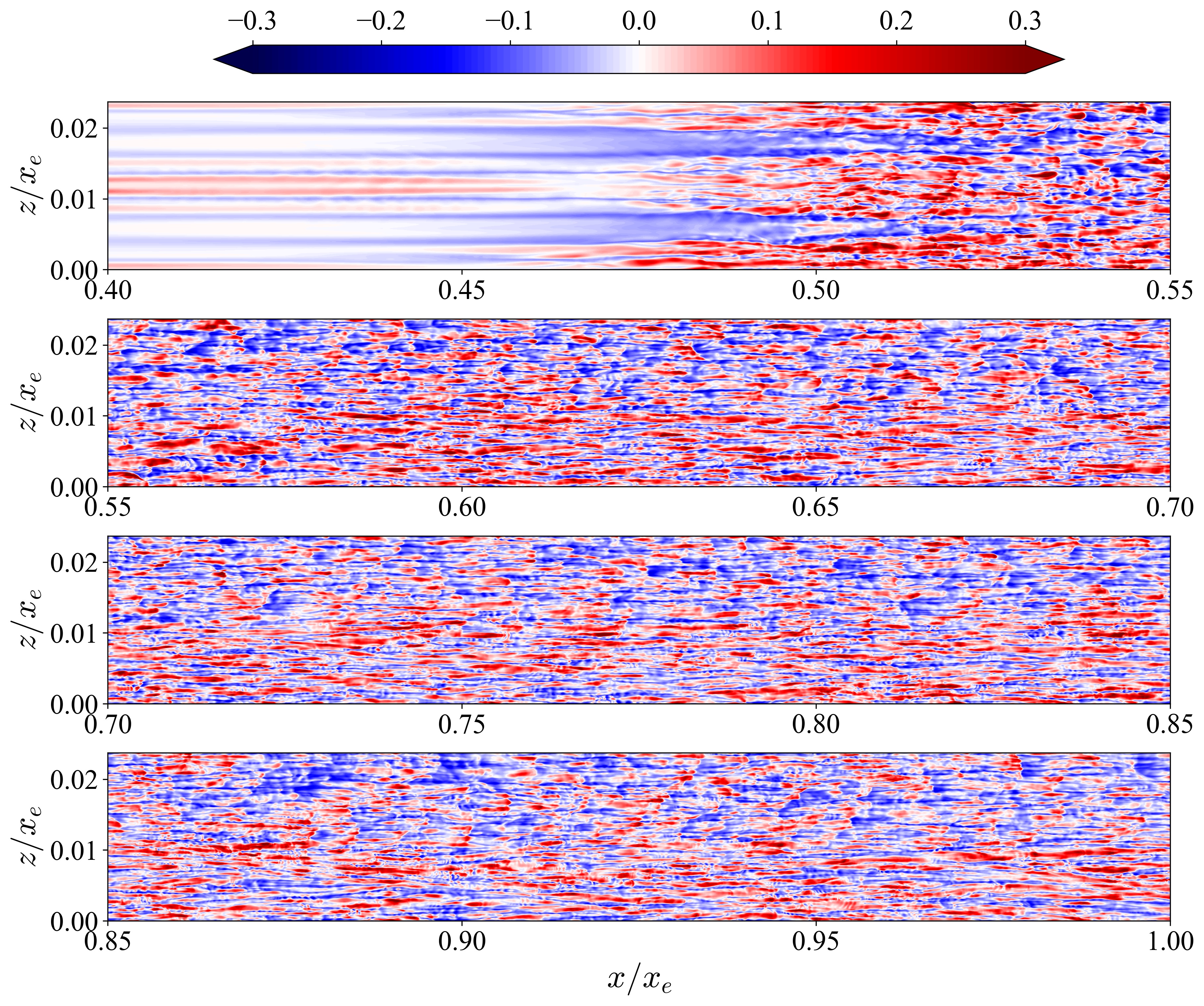}
	\caption{Instantaneous field of ${u}''$ in a wall-parallel plane at $y^{+}=10$ in M8T015.}
	\label{fig: d2-1}
\end{figure}

\begin{figure}\centering
         \includegraphics[width=0.96\linewidth]{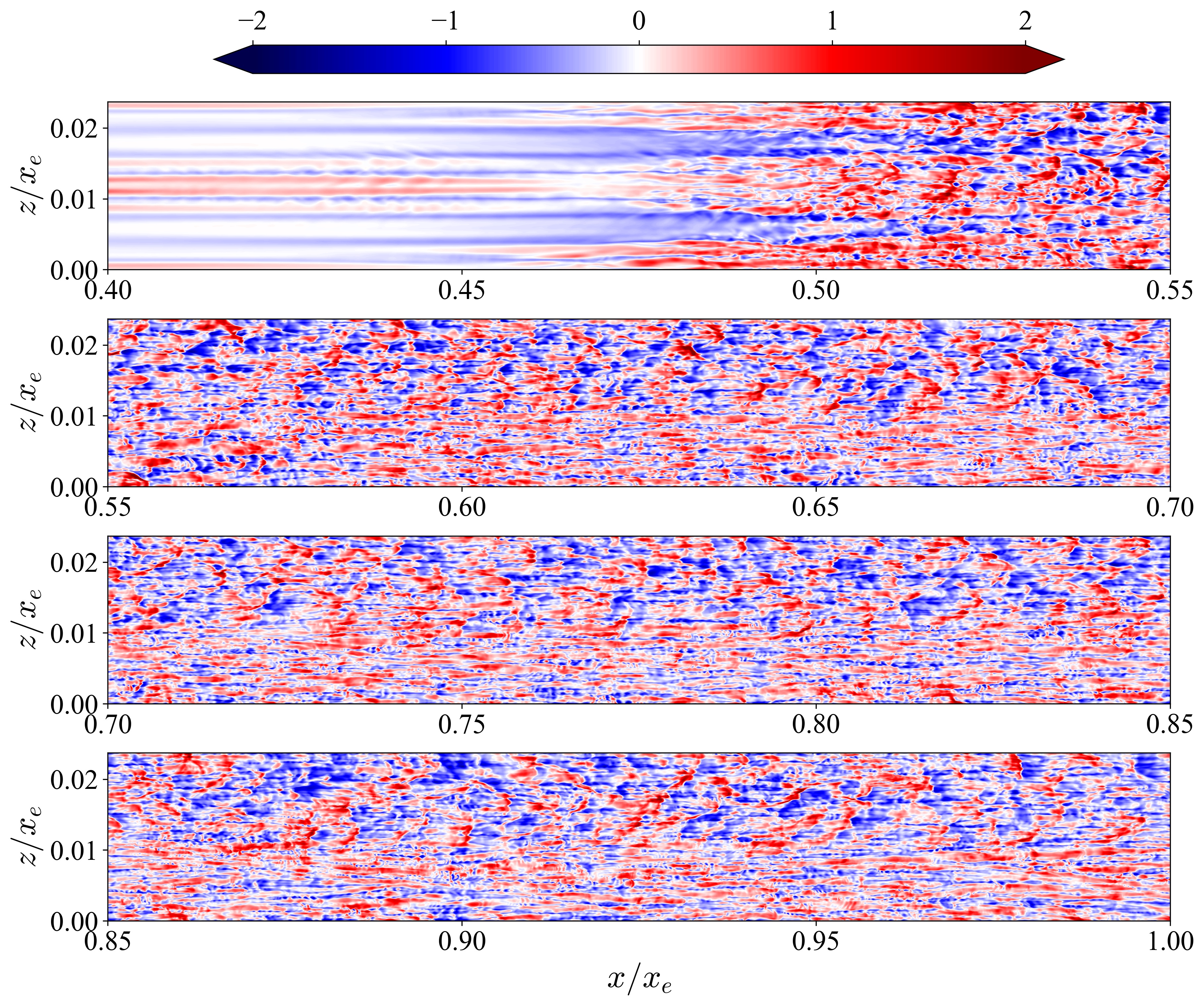}
	\caption{Instantaneous field of ${T}''$ in a wall-parallel plane at $y^{+}=10$ in M8T015.}
	\label{fig: d2-2}
\end{figure}

Several hypersonic transitional and turbulent boundary layers at Mach numbers 6 and 8 with different wall temperatures are investigated numerically in the present study. The fundamental parameters of the database are listed in table \ref{tab: tab2}. Here $M_{\infty }= U_{\infty }/c_{\infty }$ and $Re_{\infty }= \rho _{\infty }U_{\infty }L_{\infty }/\mu _{\infty }$ are the free-stream Mach number and Reynolds number respectively. The free-stream temperature $T_{\infty }$ is assumed to be $T_{\infty }=169.44K$. Temperature $T_{w}$ is the wall temperature and the recovery temperature $T_{r}$ is defined as $T_{r}=T_{\infty }\left ( 1+r\left ( \left ( \gamma -1 \right )/2 \right )M_{\infty }^{2} \right )$ with recovery factor $r=0.9$ \citep[]{Duan2010}. $x$, $y$ and $z$ are three coordinates along the streamwise, wall-normal and spanwise directions respectively. $N_{x}$, $N_{y}$ and $N_{z}$ are the grid resolutions along the streamwise, wall-normal and spanwise directions respectively. $\Delta x^{+}=\Delta x /\delta _{\nu}$, $\Delta y_{w}^{+}=\Delta y_{w} /\delta _{\nu}$ and $\Delta z^{+}=\Delta z /\delta _{\nu}$ are the normalized spacings of streamwise direction, first point off the wall and spanwise direction respectively, where $\delta _{\nu}$ is the viscous length scale. The ending streamwise location of the fully developed region shown in figure \ref{fig: d1} (a) is represented as $x_{e}$ and listed in table \ref{tab: tab2}. $N$ flow-field snapshots spanning a time interval $T_{f}$ are used for further analysis. It is noted that the adequacy of the computational domain size in the spanwise direction and the grid convergence studies of the hypersonic transitional and turbulent boundary layer database have been confirmed in \citet[]{Xu2021,Xu2021b} previously.

\begin{table}
	\begin{center}
		\def~{\hphantom{0}}
		\begin{tabular}{ccccccc}
			Case & $M_{\infty }$ &$Re_{\infty }$ &$T_{w}/T_{\infty }$& $T_{w}/T_{r}$ &$L_{x}\times L_{y}\times L_{z}$ &$N_{x}\times N_{y}\times N_{z}$ \\ [3pt]
            M6T04 & 6 &$2\times 10^{6}$ & 3.0 &0.4 & $16 \times 0.7 \times 0.2$ &$7000\times200\times640$ \\
			M6T08 & 6 &$2\times 10^{6}$ & 6.0 & 0.8 & $16 \times 0.7 \times 0.2$ &$6000\times200\times320$ \\
			M8T015 & 8 &$2\times 10^{6}$ & 1.9 &0.15& $19\times 0.7\times 0.35$ & $9000\times200\times1280$ \\
			M8T04 & 8 &$2\times 10^{6}$ & 5.0 & 0.4 & $17\times 0.7\times 0.2$ & $7000\times200\times320$ \\
			M8T08 &8 &$5\times 10^{6}$ & 10.03 &0.8 & $41 \times 0.7 \times 0.6$ &$12500\times200\times640$ \\
		\end{tabular}
\setlength{\tabcolsep}{5mm}
		\begin{tabular}{ccccccc}
			Case &  $\Delta x^{+}$ & $\Delta y_{w}^{+}$ &  $\Delta z^{+}$ & $x_{e}$ & $N$ & $T_{f}$\\
            M6T04  &  9.3 & 0.5 & 3.6 & 13 &  300 & $5x_{e} /U_{\infty }$ \\
            M6T08  &  4.9 & 0.5 & 3.1 & 13 &  300 & $5x_{e} /U_{\infty }$\\
			M8T015 &  11.2 & 0.5 & 4.5 & 15 &  300 & $5x_{e} /U_{\infty }$\\
            M8T04  &  5.6 & 0.5 & 3.5 & 14 &  300 & $5x_{e} /U_{\infty }$\\
			M8T08  & 12.2 & 0.5 & 4.6 & 34 &  300 & $5x_{e} /U_{\infty }$\\ 			
		\end{tabular}
		\caption{Summary of computational parameters for the DNS study. The computational domains $L_{x}$, $L_{y}$ and $L_{z}$ are nondimensionalized by 1 \emph{inch} \citep[]{Pirozzoli2004}.}
		\label{tab: tab2}
	\end{center}
\end{table}

The Reynolds analogy factor $s$ represents the ratio between the wall heat transfer and the wall skin friction, and can be defined as \citep[]{Zhang2014,Wenzel2021a}
\begin{equation}
s\equiv \frac{2C_{h}}{C_{f}}=\frac{\overline{q}_{w}u_{\delta }}{\overline{\tau }_{w}C_{p}\left ( T_{w}-T_{r} \right )}.		
\end{equation}
Here $C_{f}\equiv \overline{\tau }_{w}/\left ( \overline{\rho} _{\delta }\overline{u}_{\delta }^{2}/2 \right )$ is the skin-friction coefficient, and $C_{h}\equiv \overline{q}_{w}/\left ( \overline{\rho} _{\delta }\overline{u}_{\delta }C_{p}\left ( T_{w}-T_{r} \right ) \right )$ is the heat transfer coefficient, i.e. the Stanton number. The subscript $\delta $ represents the flow variables evaluated at the boundary-layer edge. $\overline{\tau }_{w}\equiv  \left (\mu \partial \overline{u}/\partial y  \right ) _{y=0}$ is the mean wall shear stress, and $\overline{q}_{w}\equiv -\left ( \overline{\left ( \kappa /\alpha  \right )\left (\partial T/\partial y  \right )} \right )_{y=0}$ is the wall heat flux. The Reynolds analogy factor $s$ reveals the integral effect of the momentum and heat transfer across the boundary layer at the wall. The Reynolds analogy factor $s$ along the streamwise direction in the hypersonic transitional and turbulent boundary layers is shown in figure~\ref{fig: d1} (b). It is found that the Reynolds analogy factor $s$ is relatively small in the laminar state, mainly due to the reason that conduction is the primary heat transfer process in this state and has a low heat transfer efficiency. Then, the laminar flow undergoes the laminar-to-turbulent bypass transition, and the Reynolds analogy factor $s$ increases. An overshoot phenomenon is also found at the transition peak, which is similar to the behaviours of the skin-friction and heat transfer coefficients in many previous studies  \citep[]{Pirozzoli2004,Franko2011,Franko2013,Liang2013,Liang2015,Xu2021b}. After the transition peak, the Reynolds analogy factor $s$ decreases to the turbulent values, and then remains nearly constant in the fully developed region, indicating that the relatively weak Reynolds-number dependency of $s$ in the fully developed region for the moderate Reynolds-number range considered here \citep[]{Wenzel2021a}. It is also found that the Reynolds analogy factor $s$ of the turbulent flows is much larger than that of the laminar flows, due to the reason that turbulent mixing is the dominant mechanism of momentum and heat transfer in turbulent boundary layers, and convection becomes the dominant heat transfer process and has a high heat transfer efficiency. For compressible turbulent boundary layers, the behaviour of the Reynolds analogy factor $s$ is complex and can be relevant to Mach number, wall temperature, surface roughness, pressure gradients and so on. Previous observations indicated that $s$ is insensitive to variations of Mach number and wall temperature \citep[]{Duan2010,Zhang2014,Wenzel2021a}, while surface roughness \citep[]{Peeters2019} and pressure gradient \citep[]{Wenzel2021a} have strong effects on $s$. However, as shown in figure \ref{fig: d1} (b), the Reynolds analogy factor $s$ slightly increases as the wall temperature decreases, especially for the extremely cold wall temperature case ``M8T015'', which has the largest values of $s$. It is should be noted that the influence of the wall temperature is much weak compared with those of the surface roughness and pressure gradient.

The visualization of the instantaneous vortical structures based on the Q-criterion \citep[]{Hunt1988} of the hypersonic transitional and turbulent boundary layer in M8T015 is shown in figure \ref{fig: d2}. The structures are coloured by the wall-normal location $y/\delta$. This picture depicts the streamwise evolution of the instantaneous vortical structures in the whole laminar-to-turbulent transition process. It is found that at the onset of the flat plate, the laminar flow is disturbed by the wall blowing and suction, and then results in the streamwise vorticity near the wall (appears at the streamwise position $x/x_{e}=0.4-0.45$ in figure \ref{fig: d2}). Furthermore, the high- and low- momentum streaks are generated by the lift-up effect \citep[]{Franko2013}, and more and more vortical structures appear far from the wall as the boundary layer evolves. It is also found that the vortical structures far from the wall are blob-like structures, while those near the wall are sheet-like structures, mainly due to the strong mean shear in the near-wall region.

Furthermore, instantaneous fields of ${u}''$ and ${T}''$ in a wall-parallel plane at $y^{+}=10$ in M8T015 are plotted in figures \ref{fig: d2-1} and \ref{fig: d2-2} respectively. It is found that the structures of ${u}''$ and ${T}''$ are similar to each other, which is consistent with the observation in \citet[]{Xu2021} that the correlation values between ${u}''$ and ${T}''$ are almost equal to 1 in the near-wall region. At the onset of the bypass transition process (at the streamwise position $x/x_{e}=0.4-0.45$), the streak instability shows to be a sinuous structure, which is consistent with the behaviours in compressible \citep[]{Franko2013} and incompressible \citep[]{Andersson2001} flows. The sinuous disturbances lead to the high- and low- momentum and temperature streaks, and these streaks then break down to a fully turbulent flows \citep[]{Franko2013}.

\section{Decompositions of the skin-friction and heat transfer coefficients}\label{sec: n2}
Integral identities are helpful to investigate the wall skin friction and heat transfer by the contributions from the flow statistics inside the flow domain, which can help to figure out the generation mechanisms of wall skin friction and heat transfer and design flow-control strategies. The integral analysis has been used widely in previous investigations by integrating different types of conservation laws, including mean momentum \citep[]{Fukagata2002,Xia2021,Wenzel2021b}, mean kinetic energy \citep[]{Renard2016}, mean energy \citep[]{Arntz2015} and mean internal energy \citep[]{Zhang2020}. In this study, the integral identities of the mean momentum and internal energy are analyzed to investigate the contributions of internal flow statistics to the skin-friction coefficient $C_{f}$ as well as the heat transfer coefficient $C_{h}$.

The governing equation of the internal energy can be expressed as
\begin{equation}
	\frac{\partial \left (\rho C_{v} T  \right )}{\partial t}+\frac{\partial \left (\rho C_{v} Tu_{j}  \right )}{\partial x_{j}}=-\rho \left ( \gamma -1 \right )C_{v}T\frac{\partial u_{j}}{\partial x_{j}}+\frac{\sigma _{ij}}{Re}\frac{\partial u_{i}}{\partial x_{j}}+\frac{1}{\alpha }\frac{\partial }{\partial x_{j}}\left ( \kappa \frac{\partial T}{\partial x_{j}} \right ).
\end{equation}

Accordingly, the stationary Reynolds averages of density, momentum and internal energy equations are defined as
\begin{equation}
\frac{\partial \left (\bar{\rho }\widetilde{u}_{j}  \right )}{\partial x_{j}}=0,
\end{equation}
\begin{equation}
	\frac{\partial \left (\bar{\rho }\widetilde{u}_{i}\widetilde{u}_{j}  \right )}{\partial x_{j}}+\frac{\partial \left ( \overline{\rho {u}''_{i}{u}''_{j}} \right )}{\partial x_{j}}=-\frac{\partial \bar{p}\delta _{ij}}{\partial x_{j}}+\frac{\partial \overline{\sigma }_{ij}}{\partial x_{j}},
\end{equation}
\begin{equation}
	\frac{\partial \left (\bar{\rho} \widetilde{T}\widetilde{u}_{j}  \right )}{\partial x_{j}}+\frac{\partial \left ( \overline{\rho {T}''{u}''_{j}} \right )}{\partial x_{j}}=-\overline{\rho \left ( \gamma -1 \right )T\frac{\partial u_{j}}{\partial x_{j}}}+\overline{\frac{\sigma _{ij}}{C_{v}Re}\frac{\partial u_{i}}{\partial x_{j}}}+\overline{\frac{1}{C_{v}\alpha }\frac{\partial }{\partial x_{j}}\left ( \kappa \frac{\partial T}{\partial x_{j}} \right )}.
\end{equation}
Then, the 2D-boundary-layer equations for the mean streamwise momentum and mean temperature can be written as
\begin{equation}
	\begin{aligned}
	&-\bar{\rho }\widetilde{u}\frac{\partial \widetilde{u}}{\partial x}-\bar{\rho }\widetilde{v}\frac{\partial \widetilde{u}}{\partial y}-\frac{\partial \bar{p}}{\partial x}+\frac{1}{Re}\left ( \frac{\partial \overline{\sigma}_{xx} }{\partial x}
	+ \frac{\partial \overline{\sigma}_{xy} }{\partial y}\right )\\
	&-\frac{\partial \left (\bar{\rho }\widetilde{{u}''{u}''} \right )}{\partial x}-\frac{\partial \left (\bar{\rho }\widetilde{{u}''{v}''}\right )}{\partial y}=0,
\end{aligned}
\label{eqn: xmom}
\end{equation}
\begin{equation}
\begin{aligned}
	&-\bar{\rho }\widetilde{u}\frac{\partial \widetilde{T}}{\partial x}-\bar{\rho }\widetilde{v}\frac{\partial \widetilde{T}}{\partial y}-\overline{\rho \left ( \gamma -1 \right )T\frac{\partial u_{j}}{\partial x_{j}}}+\overline{\frac{\sigma _{ij}}{C_{v}Re}\frac{\partial u_{i}}{\partial x_{j}}}+\overline{\frac{1}{C_{v}\alpha }\frac{\partial }{\partial x}\left ( \kappa \frac{\partial T}{\partial x} \right )} \\
	&+\overline{\frac{1}{C_{v}\alpha }\frac{\partial }{\partial y}\left ( \kappa \frac{\partial T}{\partial y} \right )}-\frac{\partial \left (\bar{\rho }\widetilde{{T}''{u}''} \right )}{\partial x}-\frac{\partial \left (\bar{\rho }\widetilde{{T}''{v}''}\right )}{\partial y}=0.
\end{aligned}
\label{eqn: Tm}
\end{equation}

The Cauchy$^{'}$s formula indicates that $n$ repeated integrations of a continuous function can be expressed as a single integral \citep[]{Wenzel2021b}
\begin{equation}
	\int_{y_{l}}^{y_{b}}\int_{y_{l}}^{y_{2}}\cdots \int_{y_{l}}^{y_{n-1}}f\left (y_{n}  \right )dy_{n}\cdots dy_{2}dy=\frac{1}{\left ( n-1 \right )!}\int_{y_{l}}^{y_{b}}\left ( y_{b}-y \right )^{n-1}f\left ( y \right )dy. \label{eqn: Cauchy}
\end{equation}
If $n$ repeated integration is applied on equations \ref{eqn: xmom} and \ref{eqn: Tm}, according to the Cauchy$^{'}$s formula equation \ref{eqn: Cauchy}, the skin-friction coefficient $C_{f}$ and heat transfer coefficient $C_{h}$ can be expressed as
\begin{equation}
\begin{small}
	\begin{aligned}
		C_{f}&=\frac{2\left ( n-1 \right )}{\overline{\rho} _{\delta }\overline{u}_{\delta }^{2}y_{b}^{n-1}}\int_{0}^{y_{b}}\left ( y_{b}-y \right )^{n-2}\left [ \frac{1}{Re}\left ( \bar{\mu }\frac{\partial \bar{u}}{\partial y}+\bar{\mu }\frac{\partial \bar{v}}{\partial x}+ \overline{{\mu }'\left ( \frac{\partial {u}'}{\partial y}+\frac{\partial {v}'}{\partial x} \right )}\right )-\bar{\rho }\widetilde{{u}''{v}''} \right ]dy \\
		&+\frac{2}{\overline{\rho} _{\delta }\overline{u}_{\delta }^{2}y_{b}^{n-1}}\int_{0}^{y_{b}}\left ( y_{b}-y \right )^{n-1}\left [ -\bar{\rho }\widetilde{u}\frac{\partial \widetilde{u}}{\partial x}-\bar{\rho }\widetilde{v}\frac{\partial \widetilde{u}}{\partial y}-\frac{\partial \bar{p}}{\partial x} +\frac{1}{Re}\frac{\partial \overline{\sigma}_{xx} }{\partial x}-\frac{\partial \left (\bar{\rho }\widetilde{{u}''{u}''} \right )}{\partial x}\right ]dy,
\end{aligned}
\end{small}
\label{eqn: Cfa}
\end{equation}
\begin{equation}
\begin{small}
	\begin{aligned}
		C_{h}&=\frac{C_{v}\left ( n-1 \right )}{\left ( \overline{\rho} _{\delta }\overline{u}_{\delta }C_{p}\left ( T_{r}-T_{w} \right ) \right )y_{b}^{n-1}}\int_{0}^{y_{b}}\left ( y_{b}-y \right )^{n-2}\left [\overline{\frac{1}{C_{v}\alpha }\left ( \kappa \frac{\partial T}{\partial y} \right )}-\bar{\rho }\widetilde{{T}''{v}''}  \right ]dy \\
		&+\frac{C_{v}}{\left ( \overline{\rho} _{\delta }\overline{u}_{\delta }C_{p}\left ( T_{r}-T_{w} \right ) \right )y_{b}^{n-1}}\int_{0}^{y_{b}}\left ( y_{b}-y \right )^{n-1}\left [-\bar{\rho }\widetilde{u}\frac{\partial \widetilde{T}}{\partial x}-\bar{\rho }\widetilde{v}\frac{\partial \widetilde{T}}{\partial y} \right ]dy \\
&+\frac{C_{v}}{\left ( \overline{\rho} _{\delta }\overline{u}_{\delta }C_{p}\left ( T_{r}-T_{w} \right ) \right )y_{b}^{n-1}}\int_{0}^{y_{b}}\left ( y_{b}-y \right )^{n-1}\left [-\frac{\partial \left (\bar{\rho }\widetilde{{T}''{u}''} \right )}{\partial x}-\overline{\rho \left ( \gamma -1 \right )T\frac{\partial u_{j}}{\partial x_{j}}} \right ]dy \\
		& +\frac{C_{v}}{\left ( \overline{\rho} _{\delta }\overline{u}_{\delta }C_{p}\left ( T_{r}-T_{w} \right ) \right )y_{b}^{n-1}}\int_{0}^{y_{b}}\left ( y_{b}-y \right )^{n-1}\left [\overline{\frac{\sigma _{ij}}{C_{v}Re}\frac{\partial u_{i}}{\partial x_{j}}}+\overline{\frac{1}{C_{v}\alpha }\frac{\partial }{\partial x}\left ( \kappa \frac{\partial T}{\partial x} \right )}  \right ]dy,
	\end{aligned}
\end{small}
\label{eqn: Cha}
\end{equation}
where $y_{b}$ is the upper integration limit. It should be noted that an arbitrary values of $n$ $\left ( n>1 \right )$ can lead to a valid decomposition of $C_{f}$ and $C_{h}$; however, not all values of $n$ can result in a physically plausible interpretation \citep[]{Wenzel2021b}. The well-known FIK identity \citep[]{Fukagata2002} is based on a three-fold $\left ( n=3 \right )$ integration of the streamwise momentum equation, but a drawback of the FIK identity is the lack of a physical-based interpretation of some contributing terms \citep[]{Renard2016,Li2019,Wenzel2021b}. As suggested by \citet[]{Ghosh2010}, \citet[]{Zhang2020} and \citet[]{Wenzel2021b}, a two-fold $\left ( n=2 \right )$ repeated integration is more suitable. The reason can be explained as follows. The first integration leads to a force and energy balance between the wall and other wall-normal locations within the boundary layer, and the second integration leads to an average in the wall-normal direction \citep[]{Zhang2020,Wenzel2021b}. Accordingly, if $n=2$, the equations \ref{eqn: Cfa} and \ref{eqn: Cha} can be simplified as
\begin{equation}
C_{f}=C^{B}_{f}+C^{V}_{f}+C^{T}_{f}+C^{M}_{f}+C^{D,1}_{f}+C^{D,2}_{f}+C^{D,3}_{f}+C^{D,4}_{f}, \label{eqn: Cf}
\end{equation}
\begin{equation}
C_{h}=C^{B}_{h}+C^{T}_{h}+C^{V}_{h}+C^{M}_{h}+C^{P}_{h}+C^{D,1}_{h}+C^{D,2}_{h}+C^{D,3}_{h}, \label{eqn: Ch}
\end{equation}
where
\begin{equation}
C_{f}^{B}\equiv \frac{2}{\overline{\rho} _{\delta }\overline{u}_{\delta }^{2}y_{b}}\int_{0}^{y_{b}}\left [ \frac{1}{Re}\left ( \bar{\mu }\frac{\partial \bar{u}}{\partial y}+\bar{\mu }\frac{\partial \bar{v}}{\partial x} \right ) \right ]dy,
\end{equation}
\begin{equation}
C_{f}^{V}\equiv \frac{2}{\overline{\rho} _{\delta }\overline{u}_{\delta }^{2}y_{b}}\int_{0}^{y_{b}}\left [ \overline{{\mu }'\left ( \frac{\partial {u}'}{\partial y}+\frac{\partial {v}'}{\partial x} \right )} \right ]dy,
\end{equation}
\begin{equation}
C_{f}^{T}\equiv -\frac{2}{\overline{\rho} _{\delta }\overline{u}_{\delta }^{2}y_{b}}\int_{0}^{y_{b}}\bar{\rho }\widetilde{{u}''{v}''}dy,
\end{equation}
\begin{equation}
C_{f}^{M}\equiv -\frac{2}{\overline{\rho} _{\delta }\overline{u}_{\delta }^{2}y_{b}}\int_{0}^{y_{b}}\left ( y_{b}-y \right )\bar{\rho }\widetilde{v}\frac{\partial \widetilde{u}}{\partial y}dy,
\end{equation}
\begin{equation}
C_{f}^{D,1}\equiv -\frac{2}{\overline{\rho} _{\delta }\overline{u}_{\delta }^{2}y_{b}}\int_{0}^{y_{b}}\left ( y_{b}-y \right ) \bar{\rho }\widetilde{u}\frac{\partial \widetilde{u}}{\partial x}dy,
\end{equation}
\begin{equation}
C_{f}^{D,2}\equiv -\frac{2}{\overline{\rho} _{\delta }\overline{u}_{\delta }^{2}y_{b}}\int_{0}^{y_{b}}\left ( y_{b}-y \right ) \frac{\partial \bar{p}}{\partial x}dy,
\end{equation}
\begin{equation}
C_{f}^{D,3}\equiv \frac{2}{\overline{\rho} _{\delta }\overline{u}_{\delta }^{2}y_{b}}\int_{0}^{y_{b}}\left ( y_{b}-y \right ) \frac{1}{Re}\frac{\partial \overline{\sigma}_{xx} }{\partial x}dy,
\end{equation}
\begin{equation}
C_{f}^{D,4}\equiv -\frac{2}{\overline{\rho} _{\delta }\overline{u}_{\delta }^{2}y_{b}}\int_{0}^{y_{b}}\left ( y_{b}-y \right )\frac{\partial \left (\bar{\rho }\widetilde{{u}''{u}''} \right )}{\partial x}dy;
\end{equation}
and
\begin{equation}
C^{B}_{h}\equiv \frac{C_{v}}{\left ( \overline{\rho} _{\delta }\overline{u}_{\delta }C_{p}\left ( T_{r}-T_{w} \right ) \right )y_{b}}\int_{0}^{y_{b}}\overline{\frac{1}{C_{v}\alpha }\left ( \kappa \frac{\partial T}{\partial y} \right )}dy,
\end{equation}
\begin{equation}
C^{T}_{h}\equiv -\frac{C_{v}}{\left ( \overline{\rho} _{\delta }\overline{u}_{\delta }C_{p}\left ( T_{r}-T_{w} \right ) \right )y_{b}}\int_{0}^{y_{b}}\bar{\rho }\widetilde{{T}''{v}''}dy,
\end{equation}
\begin{equation}
C^{V}_{h}\equiv \frac{C_{v}}{\left ( \overline{\rho} _{\delta }\overline{u}_{\delta }C_{p}\left ( T_{r}-T_{w} \right ) \right )y_{b}}\int_{0}^{y_{b}}\left ( y_{b}-y \right )\overline{\frac{\sigma _{ij}}{C_{v}Re}\frac{\partial u_{i}}{\partial x_{j}}} dy,
\end{equation}
\begin{equation}
C^{M}_{h}\equiv -\frac{C_{v}}{\left ( \overline{\rho} _{\delta }\overline{u}_{\delta }C_{p}\left ( T_{r}-T_{w} \right ) \right )y_{b}}\int_{0}^{y_{b}}\left ( y_{b}-y \right )\bar{\rho }\widetilde{v}\frac{\partial \widetilde{T}}{\partial y} dy,
\end{equation}
\begin{equation}
C^{P}_{h}\equiv -\frac{C_{v}}{\left ( \overline{\rho} _{\delta }\overline{u}_{\delta }C_{p}\left ( T_{r}-T_{w} \right ) \right )y_{b}}\int_{0}^{y_{b}}\left ( y_{b}-y \right )\overline{\rho \left ( \gamma -1 \right )T\frac{\partial u_{j}}{\partial x_{j}}} dy,
\label{eqn: Cph}
\end{equation}
\begin{equation}
C^{D,1}_{h}\equiv -\frac{C_{v}}{\left ( \overline{\rho} _{\delta }\overline{u}_{\delta }C_{p}\left ( T_{r}-T_{w} \right ) \right )y_{b}}\int_{0}^{y_{b}}\left ( y_{b}-y \right )\bar{\rho }\widetilde{u}\frac{\partial \widetilde{T}}{\partial x}dy,
\end{equation}
\begin{equation}
C^{D,2}_{h}\equiv -\frac{C_{v}}{\left ( \overline{\rho} _{\delta }\overline{u}_{\delta }C_{p}\left ( T_{r}-T_{w} \right ) \right )y_{b}}\int_{0}^{y_{b}}\left ( y_{b}-y \right )\frac{\partial \left (\bar{\rho }\widetilde{{T}''{u}''} \right )}{\partial x}dy,
\end{equation}
\begin{equation}
C^{D,3}_{h}\equiv \frac{C_{v}}{\left ( \overline{\rho} _{\delta }\overline{u}_{\delta }C_{p}\left ( T_{r}-T_{w} \right ) \right )y_{b}}\int_{0}^{y_{b}}\left ( y_{b}-y \right )\overline{\frac{1}{C_{v}\alpha }\frac{\partial }{\partial x}\left ( \kappa \frac{\partial T}{\partial x} \right )}dy.
\end{equation}
Here $C_{f}^{B}$ is the mean boundary-layer term; $C_{f}^{V}$ is the viscous boundary-layer term; $C_{f}^{T}$ is the turbulent-convection term; $C_{f}^{M}$ is the mean-convection term; $C_{f}^{D,1}$, $C_{f}^{D,2}$, $C_{f}^{D,3}$ and $C_{f}^{D,4}$ are the spatial-development terms. $C^{B}_{h}$ is the mean boundary-layer term; $C^{T}_{h}$ is the turbulent-convection term; $C^{V}_{h}$ is the viscous-dissipation term; $C^{M}_{h}$ is the mean-convection term; $C^{P}_{h}$ is the pressure-dilatation term; $C^{D,1}_{h}$, $C^{D,2}_{h}$ and $C^{D,3}_{h}$ are the spatial-development terms.

The value of the upper integration limit $y_{b}$ drastically influences the relative contributions of different terms. As confirmed in \citet[]{Wenzel2021b}, the particular value of $y_{b}$ has a weak effect on the qualitative contributions of different components if $y_{b}$ approximates to the boundary-layer thickness $\delta$. Thus, in the present study, we set the upper integration limit as $y_{b}\approx 1.5\delta $ to cover the thermal boundary layer, where $\delta$ is the boundary-layer thickness of the turbulent boundary layer in the fully developed region.

\section{Streamwise evolution of $C_{f}$ in the hypersonic transitional and turbulent boundary layers}\label{sec: n3}
\begin{figure}\centering
         \includegraphics[width=0.96\linewidth]{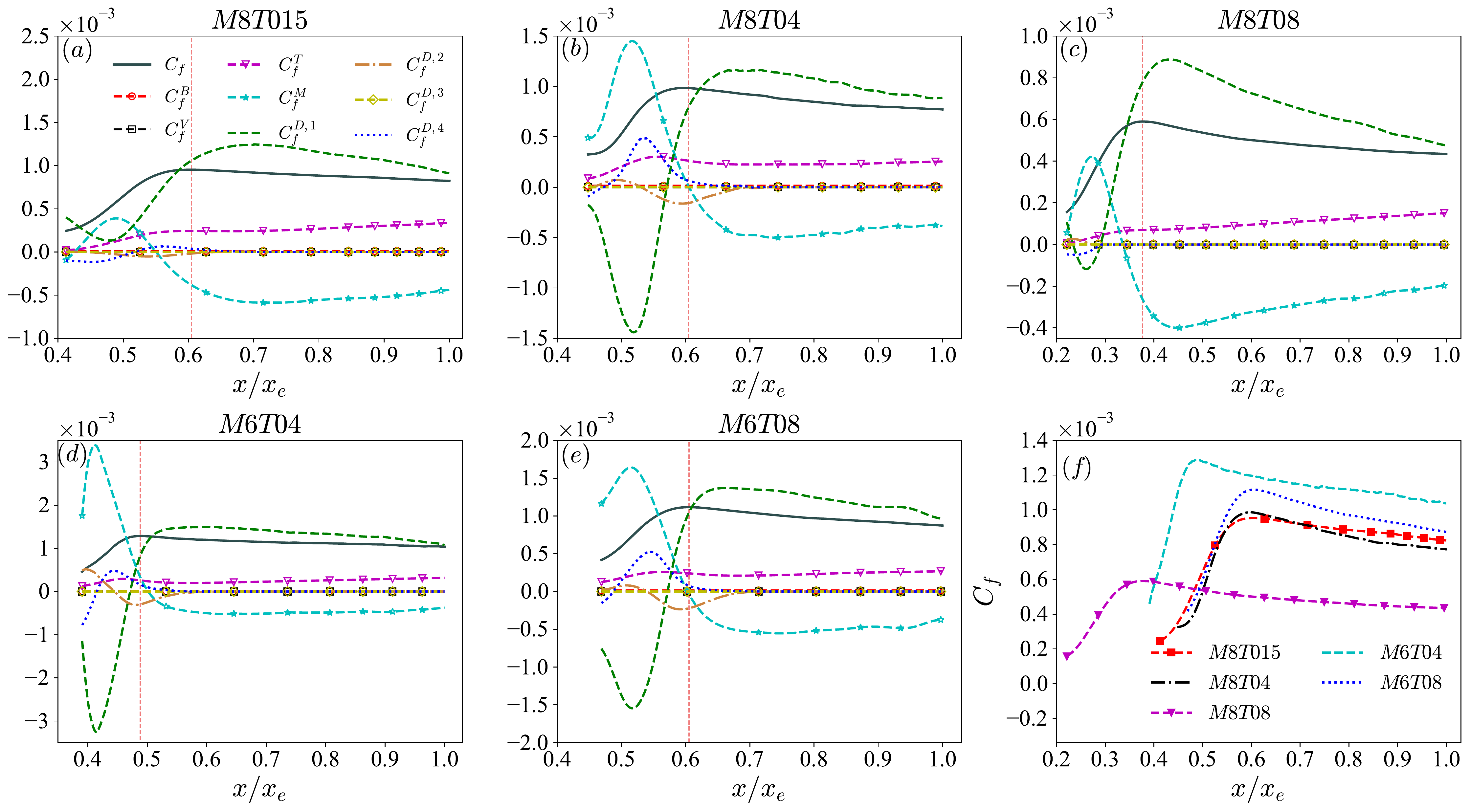}
	\caption{The streamwise evolutions of the skin-friction coefficient $C_{f}$ and its components in equation \ref{eqn: Cf} in the hypersonic transitional and turbulent boundary layers in (a) M8T015, (b) M8T04, (c) M8T08, (d) M6T04 and (e) M6T08. (f) The streamwise evolutions of the skin-friction coefficient $C_{f}$ in five hypersonic transitional and turbulent boundary layers. The vertical dashed line represents the streamwise position of the transition peak.}
	\label{fig: d3}
\end{figure}

The streamwise evolutions of the skin-friction coefficient $C_{f}$ and its components in equation \ref{eqn: Cf} in the hypersonic transitional and turbulent boundary layers are shown in figure \ref{fig: d3}. It is shown that the skin-friction coefficient $C_{f}$ initially increases due to the bypass transition process, then achieves to the peak value at the transition peak, and finally slightly decreases as the streamwise location $x/x_{e}$ increases. An interesting phenomenon is found that the peak value of the skin-friction coefficient $C_{f}$ overshoots the turbulent values in the nonlinear breakdown regime before ultimately decreases towards the turbulent boundary layer values, which has been widely found in previous investigations \citep[]{Brandt2002,Horvath2002,Pirozzoli2004,Franko2011,Franko2013,Liang2013,Liang2015}. Moreover, the values of $C_{f}$ in the fully developed region increase as the wall temperature decreases, indicating that the cold wall temperature can enhance the turbulent wall skin friction.

The trends of the contributions of the skin-friction coefficient $C_{f}$ along the streamwise direction are similar in different hypersonic transitional and turbulent boundary layers. It is found that the mean-convection term $C_{f}^{M}$ initially increases at the onset of the bypass transition process and gives the positive contribution to $C_{f}$, and then begins to decreases at the centre of the bypass transition process. After the transition peak, $C_{f}^{M}$ decreases to negative values and accordingly gives negative contribution to $C_{f}$. The magnitudes of $C_{f}^{M}$ slightly decrease as the streamwise position increases to the region of the fully developed state. The spatial-development term $C_{f}^{D,1}$ has the opposite trends along the streamwise direction compared with $C_{f}^{M}$. The terms $C_{f}^{M}$ and $C_{f}^{D,1}$ give the dominant contributions to $C_{f}$ and represent the influence of mean velocity gradients on the skin-friction coefficient. The turbulent-convection term $C_{f}^{T}$ increases along the streamwise direction and gives the positive contribution to $C_{f}$, indicating that the Reynolds shear stress is enhanced by the bypass transition process. The spatial-development terms $C_{f}^{D,2}$ and $C_{f}^{D,4}$ represent the effect of the streamwise pressure gradient and streamwise normal Reynolds stress gradient respectively. The term $C_{f}^{D,2}$ shows that the streamwise pressure gradient leads to the positive contribution at the onset of the bypass transition process, and then changes to the negative contribution to $C_{f}$ before the transition peak. After the transition peak, the term $C_{f}^{D,2}$ is almost zero due to the zero-pressure-gradient boundary layer setting. The term $C_{f}^{D,4}$ indicates that the streamwise normal Reynolds stress gradient results in the negative contribution at the initial state of the bypass transition process, and then switches to the positive contribution to $C_{f}$ before the transition peak. The term $C_{f}^{D,4}$ has negligible values after the transition peak. Other components in equation \ref{eqn: Cf} are negligible in the transitional and turbulent boundary layers.

\begin{figure}\centering
         \includegraphics[width=0.96\linewidth]{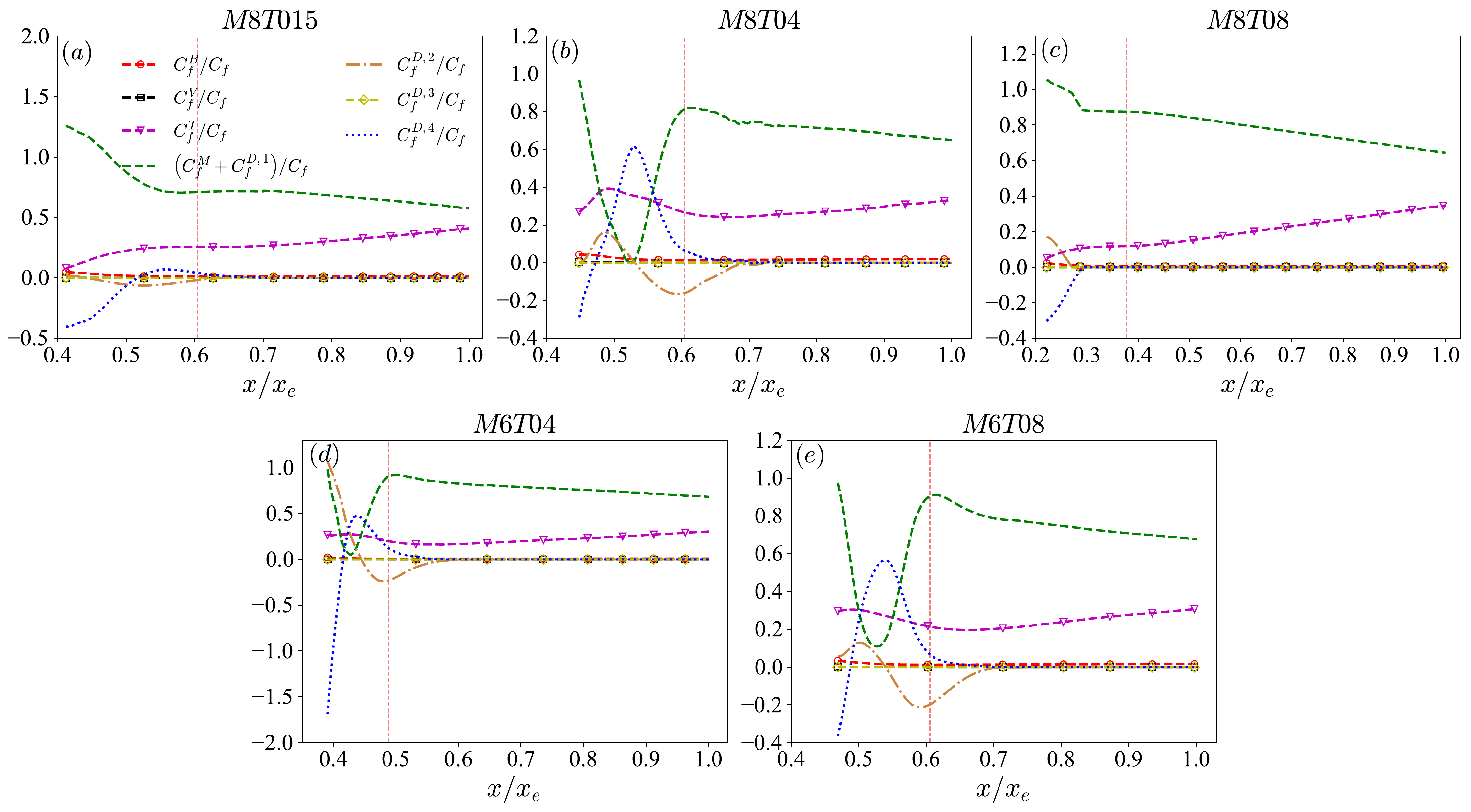}
	\caption{The streamwise evolutions of the relative contributions of the components in equation \ref{eqn: Cf} to the skin-friction coefficient $C_{f}$ in the hypersonic transitional and turbulent boundary layers in (a) M8T015, (b) M8T04, (c) M8T08, (d) M6T04 and (e) M6T08.  The vertical dashed line represents the streamwise position of the transition peak.}
	\label{fig: d4}
\end{figure}

The streamwise evolutions of the relative contributions of the components in equation \ref{eqn: Cf} to the skin-friction coefficient $C_{f}$ in the hypersonic transitional and turbulent boundary layers are plotted in figure \ref{fig: d4}. It is found that the terms $C_{f}^{M}+C_{f}^{D,1}$, $C_{f}^{T}$, $C_{f}^{D,2}$ and $C_{f}^{D,4}$ are dominant in the bypass transition process, indicating that the mean velocity gradients, the Reynolds shear stress, the streamwise pressure gradient and the streamwise normal Reynolds stress gradient are significant mechanisms during the bypass transition process. In the bypass transition process, the time-independent laminar compressible boundary-layer flow is disturbed by the wall blowing and suction, which results in the generation of the high- and low- momentum streaks by the lift-up effect \citep[]{Franko2013}. Then the streaks breakdown locally and evolve into the turbulent spots. Among the whole process, the mean velocity profiles are significantly changed by the large-amplitude fluctuations of the instability waves and result in large mean velocity gradients. Then, the breakdown of the high- and low- momentum streaks leads to the increase of the Reynolds shear stress and strong streamwise normal Reynolds stress gradient, which further results in large values of $C_{f}^{T}$ and $C_{f}^{D,4}$ in the transition process.

Furthermore, it is worth noting that the overshoot phenomenon of the skin-friction coefficient $C_{f}$ widely appears in previous experimental and numerical investigations, while its underlying mechanism is still unclear \citep[]{Franko2013}. Here the potential interpretation is given based on the integral analysis. It is shown that at the transition peak, the term $C_{f}^{M}+C_{f}^{D,1}$ gives the dominant positive contribution, and the term $C_{f}^{T}$ also gives a large positive contribution to the skin-friction coefficient $C_{f}$. Moreover, the terms $C_{f}^{D,2}$ and $C_{f}^{D,4}$ have relative small negative and positive contributions to $C_{f}$ respectively. It is noted that the value of $\left (C_{f}^{M}+C_{f}^{D,1}  \right )/C_{f}$ at the transition peak is larger than that in the fully developed state, while that of $C_{f}^{T}/C_{f}$ at the transition peak is smaller than that in the fully developed state. Therefore, the Reynolds shear stress at the transition peak is not a dominant mechanism of the overshoot phenomenon, which is different from the conclusion by \citet[]{Franko2013}; instead, the overshoot of $C_{f}$ is mainly caused by the drastic change of the mean velocity profiles.

\begin{figure}\centering
         \includegraphics[width=0.96\linewidth]{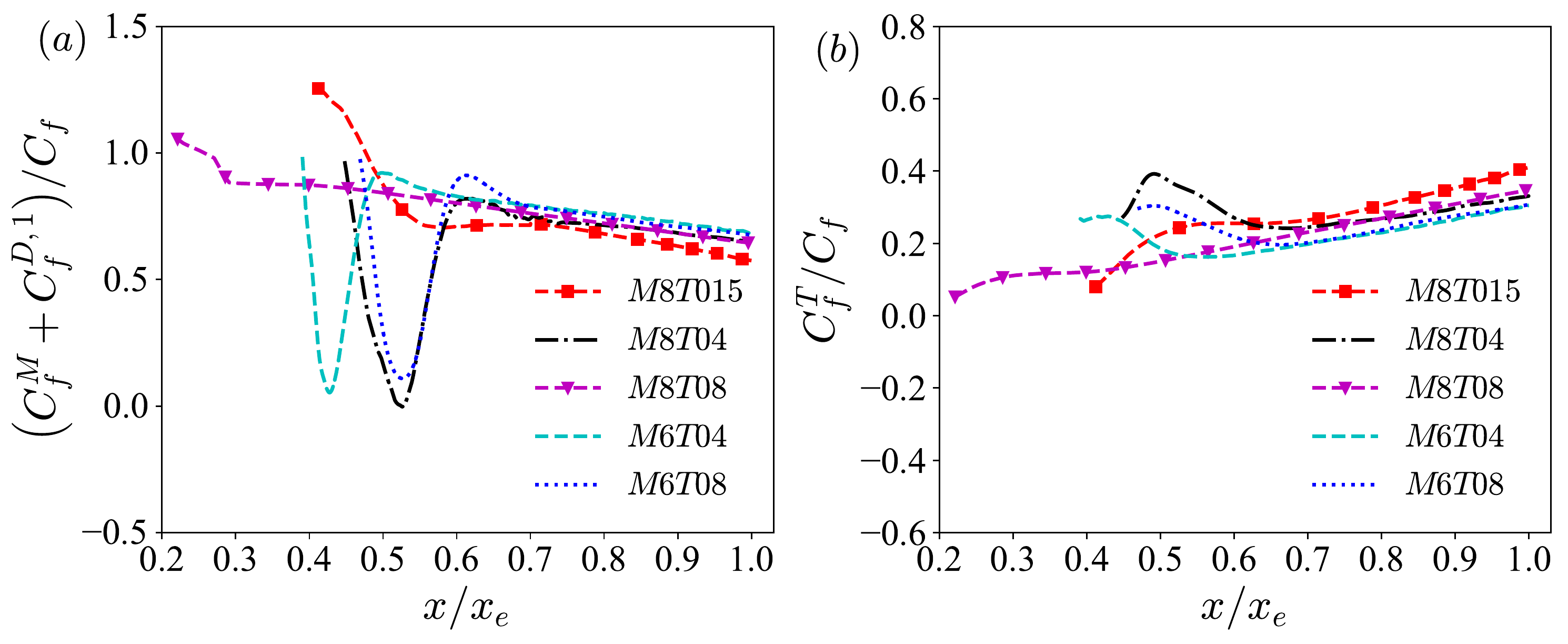}
	\caption{The streamwise evolutions of the relative contributions of (a) $C_{f}^{M}+C_{f}^{D,1}$ and (b) $C_{f}^{T}$ to the skin-friction coefficient $C_{f}$ in the hypersonic transitional and turbulent boundary layers.}
	\label{fig: d5}
\end{figure}

However, when the flow evolves to the fully developed state, only the terms $C_{f}^{M}+C_{f}^{D,1}$ and $C_{f}^{T}$ are positive, while other terms are negligibly small. It is found that the term $C_{f}^{M}+C_{f}^{D,1}$ has the dominant contribution to $C_{f}$, which is consistent with the observation in subsonic and supersonic turbulent boundary layers \citep[]{Wenzel2021b}. The relative contribution of $C_{f}^{M}+C_{f}^{D,1}$ to $C_{f}$ slightly decreases, while that of $C_{f}^{T}$ mildly increases as the streamwise position increases until to the fully developed region. The streamwise evolutions of the relative contributions of $C_{f}^{M}+C_{f}^{D,1}$ and $C_{f}^{T}$ to the skin-friction coefficient $C_{f}$ in the hypersonic transitional and turbulent boundary layers are shown in figure \ref{fig: d5}. It is found that the relative contributions of $C_{f}^{M}+C_{f}^{D,1}$ and $C_{f}^{T}$ to the skin-friction coefficient $C_{f}$ in M8T04 and M6T04 are similar to those in M8T08 and M6T08 respectively in the fully developed region, indicating that the relative contributions of the mean velocity gradients and the Reynolds shear stress to the skin-friction coefficient have weak correlations with the wall temperature in hypersonic turbulent boundary layers, which is consistent with the observation in subsonic and supersonic turbulent boundary layers \citep[]{Wenzel2021b}. This observation is satisfied in weak cooled cases (including M6T04, M6T08, M8T04 and M8T08); however, the conclusion should be modified in the strongly cooled condition, i.e. M8T015. In M8T015, the relative contribution of $C_{f}^{M}+C_{f}^{D,1}$ is smaller, while that of $C_{f}^{T}$ to the skin-friction coefficient $C_{f}$ is larger compared with those in M8T04 and M8T08, indicating that the strongly cooled wall condition can enhance the relative contribution of the Reynolds shear stress, and weaken the effect of the mean velocity gradients on the wall skin friction. It is noted that the relative contribution of each components essentially corresponds to a normalization with the wall shear stress $\bar{\tau }_{w}$. Accordingly, it is concluded in \citet[]{Wenzel2021b} that due to the Morkovin$^{'}$s scaling, the relative contribution of $C_{f}^{T}$ is Mach-number invariant. However, as the Mach number increases and the wall temperature decreases, the compressibility effect is enhanced and the Morkovin$^{'}$s scaling is not valid. Therefore, it is found that the relative contribution of $C_{f}^{T}$ increases, and that of $C_{f}^{M}+C_{f}^{D,1}$ decreases as Mach number increases, implying that the high Mach number condition can enhance the effect of the Reynolds shear stress, and weaken the impact of the mean velocity gradients.

\begin{figure}\centering
         \includegraphics[width=0.96\linewidth]{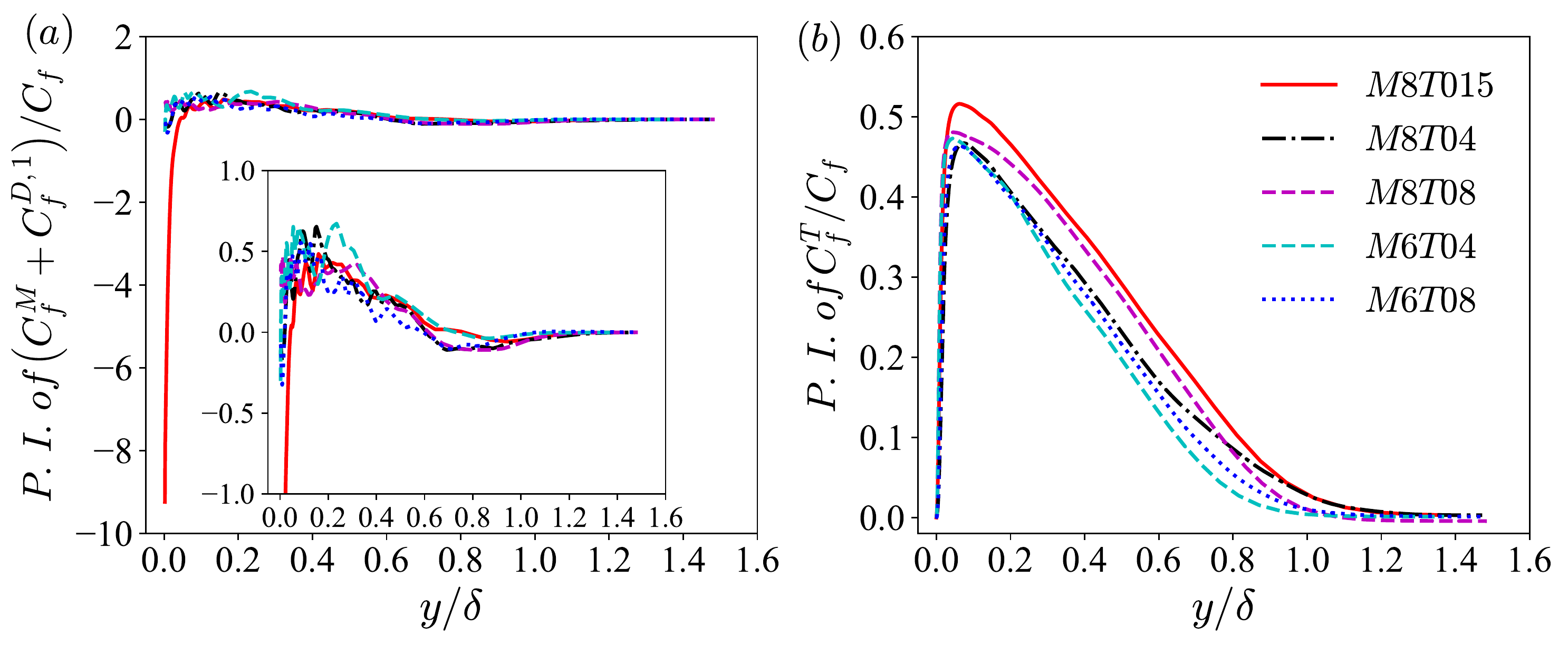}
	\caption{The pre-multiplied integrands of the relative contributions of (a) $C_{f}^{M}+C_{f}^{D,1}$ and (b) $C_{f}^{T}$ to the skin-friction coefficient $C_{f}$ in the hypersonic transitional and turbulent boundary layers at the streamwise position $x/x_{e}=0.95$. Here the abbreviation ``P. I.'' represents the pre-multiplied integrand. It is noted that the pre-multiplied integrand of $\left ( C_{f}^{M}+C_{f}^{D,1} \right )/C_{f}$ is defined as $-\frac{2}{\overline{\rho} _{\delta }\overline{u}_{\delta }^{2}C_{f}}\left ( y_{b}-y \right )\left (\bar{\rho }\widetilde{v}\frac{\partial \widetilde{u}}{\partial y}+\bar{\rho }\widetilde{u}\frac{\partial \widetilde{u}}{\partial x}  \right )$, and the pre-multiplied integrand of $C_{f}^{T}/C_{f}$ is written as $-\frac{2\bar{\rho }\widetilde{{u}''{v}''}}{\overline{\rho} _{\delta }\overline{u}_{\delta }^{2}C_{f}}$.}
	\label{fig: d6}
\end{figure}

In order to investigate the wall-normal distributions of the integrands of the components in equation \ref{eqn: Cf}, the pre-multiplied integrands of the relative contributions of $C_{f}^{M}+C_{f}^{D,1}$ and $C_{f}^{T}$ to the skin-friction coefficient $C_{f}$ in the hypersonic transitional and turbulent boundary layers at the streamwise position $x/x_{e}=0.95$ are shown in figure \ref{fig: d6}. It is found that the pre-multiplied integrands of $\left ( C_{f}^{M}+C_{f}^{D,1} \right )/C_{f}$ of all five cases are positive in most regions across the boundary layer, resulting in the positive values of $\left ( C_{f}^{M}+C_{f}^{D,1} \right )/C_{f}$. However, large negative values of the pre-multiplied integrand of $\left ( C_{f}^{M}+C_{f}^{D,1} \right )/C_{f}$ appear in the near-wall region in M8T015, which lead to smaller values of $\left ( C_{f}^{M}+C_{f}^{D,1} \right )/C_{f}$ in M8T015 compared with the other cases. Furthermore, the large negative values of the pre-multiplied integrand of $\left ( C_{f}^{M}+C_{f}^{D,1} \right )/C_{f}$ near the wall are mainly contributed by the pre-multiplied integrand of $C_{f}^{M}/C_{f}$. Accordingly, the large wall-normal gradient of the mean streamwise velocity $\partial \widetilde{u}/\partial y$ near the wall in the strongly cooled case gives rise to the smaller contribution of the mean velocity gradients to the wall skin friction.

Moreover, it is shown that the pre-multiplied integrands of $C_{f}^{T}/C_{f}$ of all cases achieve their peak values in the buffer layer and are positive along wall-normal direction, leading to the positive values of $C_{f}^{T}/C_{f}$. The peak values of the pre-multiplied integrand of $C_{f}^{T}/C_{f}$ in M8T015 are larger than those in the other cases, which gives rise to larger values of $C_{f}^{T}/C_{f}$ in M8T015. This observation indicates that the strongly cooled wall temperature enhances the normalized Reynolds shear stress, and further increases the relative contribution of the Reynolds shear stress to the skin-friction coefficient.

\section{Streamwise evolution of $C_{h}$ in the hypersonic transitional and turbulent boundary layers}\label{sec: n4}
\begin{figure}\centering
         \includegraphics[width=0.96\linewidth]{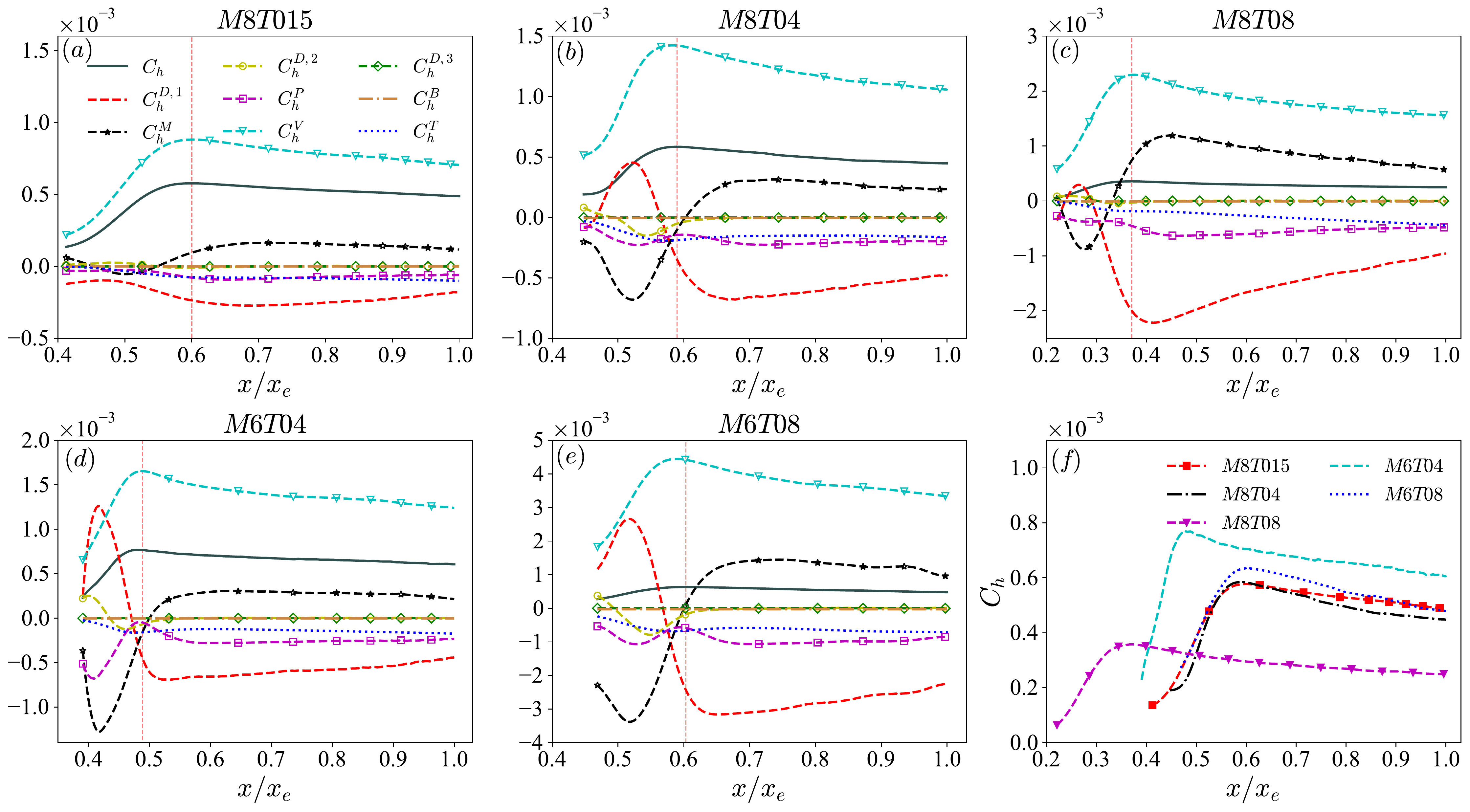}
	\caption{The streamwise evolutions of the heat transfer coefficient $C_{h}$ and its components in equation \ref{eqn: Ch} in the hypersonic transitional and turbulent boundary layers in (a) M8T015, (b) M8T04, (c) M8T08, (d) M6T04 and (e) M6T08. (f) The streamwise evolutions of the heat transfer coefficient $C_{h}$ in five hypersonic transitional and turbulent boundary layers.  The vertical dashed line represents the streamwise position of the transition peak.}
	\label{fig: d7}
\end{figure}

The streamwise evolutions of the heat transfer coefficient $C_{h}$ and its components in equation \ref{eqn: Ch} in the hypersonic transitional and turbulent boundary layers are shown in figure \ref{fig: d7}. It is found that the heat transfer coefficient $C_{h}$ has the similar trend of the skin-friction coefficient $C_{f}$ along the streamwise direction, that is, it initially increases to the transition peak due to the bypass transition process, and then slightly decreases as the streamwise position $x/x_{e}$ increases. Similar to the behaviour of the skin-friction coefficient $C_{f}$, an overshoot also appears in the streamwise evolution of $C_{h}$, which was widely observed in previous investigations \citep[]{Horvath2002,Wadhams2008,Franko2011,Franko2013}. It is also shown in figure \ref{fig: d7} (f) that the values of $C_{h}$ in the fully developed region increase significantly as the wall temperature decreases, implying that the cold wall temperature enhances the heat transfer at the wall.

The streamwise evolutions of the contributions of the heat transfer coefficient $C_{h}$ in equation \ref{eqn: Ch} along the streamwise direction in hypersonic transitional and turbulent boundary layer are similar for different Mach numbers and wall temperatures. The mean-convection term $C_{h}^{M}$ initially decreases at the incipient state of the bypass transition process, and then begins to increase at the centre of the bypass transition process. After the transition peak, $C_{h}^{M}$ increases to positive values and contributes positively to the heat transfer coefficient $C_{h}$. The spatial-development term $C_{h}^{D,1}$ has the opposite streamwise evolution compared with $C_{h}^{M}$. It is found that the terms $C_{h}^{M}$ and $C_{h}^{D,1}$ significantly contribute to the heat transfer coefficient $C_{h}$, and represent the impact of the mean temperature gradients on the heat transfer coefficient. The viscous-dissipation term $C_{h}^{V}$ increases due to the bypass transition process, and then slightly decreases after the transition peak. The magnitude of the turbulent-convection term $C_{h}^{T}$ slightly increases along the streamwise direction and gives the negative contribution to $C_{h}$, implying that the Reynolds heat flux is enhanced by the bypass transition process. The pressure-dilatation term $C_{h}^{P}$ has large negative values in the transitional and fully developed regions. The spatial-development term $C_{h}^{D,2}$ only has small values in the bypass transition process, and becomes almost zero after the transition peak, implying that the streamwise gradient of the fluctuating streamwise velocity-temperature correlation only has a limited contribution to the heat transfer coefficient in the transition process. Other terms in equation \ref{eqn: Ch} are nearly zero in the transitional and turbulent boundary layers.

\begin{figure}\centering
         \includegraphics[width=0.96\linewidth]{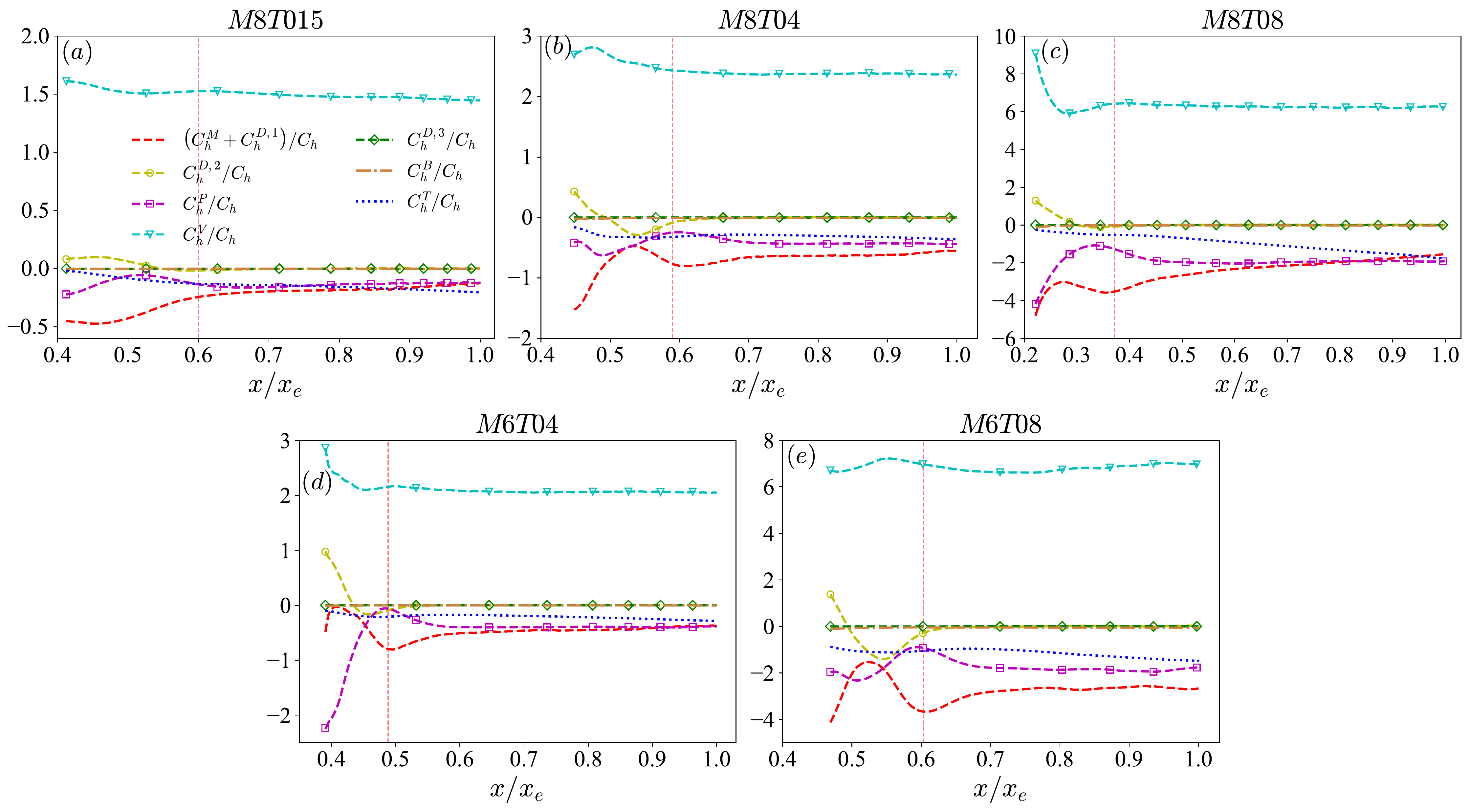}
	\caption{The streamwise evolutions of the relative contributions of the components in equation \ref{eqn: Ch} to the heat transfer coefficient $C_{h}$ in the hypersonic transitional and turbulent boundary layers in (a) M8T015, (b) M8T04, (c) M8T08, (d) M6T04 and (e) M6T08.  The vertical dashed line represents the streamwise position of the transition peak.}
	\label{fig: d8}
\end{figure}

The streamwise evolutions of the relative contributions of the components in equation \ref{eqn: Ch} to the heat transfer coefficient $C_{h}$ in the hypersonic transitional and turbulent boundary layers are shown in figure \ref{fig: d8}. It is found that the viscous-dissipation term $C_{h}^{V}$ gives the dominant positive contribution, which is consistent with the observation in supersonic turbulent channel flows by \citet[]{Zhang2020}; while the terms $C_{h}^{M}+C_{h}^{D,1}$, $C_{h}^{P}$ and $C_{h}^{T}$ give the negative contribution to the heat transfer coefficient $C_{h}$ along the transitional and turbulent boundary layers. The spatial-development term $C_{h}^{D,2}$ contributes positively at the onset of the bypass transition process, and then switches to negative at the centre of the transition process. After the transition peak, the flow evolves to the fully developed state, and the streamwise gradient of the fluctuating streamwise velocity-temperature correlation (i.e. $C_{h}^{D,2}$) decreases to almost zero.

The overshoot phenomenon of the heat transfer coefficient $C_{h}$ was extensively observed in previous investigations, while its potential mechanism is urgent to be revealed \citep[]{Franko2013}. Similar to the skin-friction coefficient $C_{f}$, an explanation to the overshoot phenomenon of the heat transfer coefficient can be given based on the integral analysis as follows. It is found that at the transition peak, the effect of viscous dissipation gives the main positive contribution and strengthens the overshoot, while the effects of the pressure dilatation, mean temperature gradients and the Reynolds heat flux contribute negatively to the heat transfer coefficient $C_{h}$ and weaken the overshoot. This observation is not consistent with the previous explanation of the overshoot of $C_{h}$ in \citet[]{Franko2013}. \citet[]{Franko2013} showed that the transport of thermal energy and the Reynolds heat flux lead to an increase in wall heat transfer and correspond to the overshoot, which is opposite to the above observation that the Reynolds heat flux weakens the overshoot phenomenon. Here, it is shown that the overshoot phenomenon of the heat transfer coefficient $C_{h}$ is mainly caused by the viscous-dissipation effect. Its underlying mechanism can be explained as follows. The large-amplitude fluctuations of the instability waves significantly change the mean velocity profiles, and the strong mean velocity gradient leads to strong viscous dissipation, which further results in the overshoot of $C_{h}$.

\begin{figure}\centering
         \includegraphics[width=0.96\linewidth]{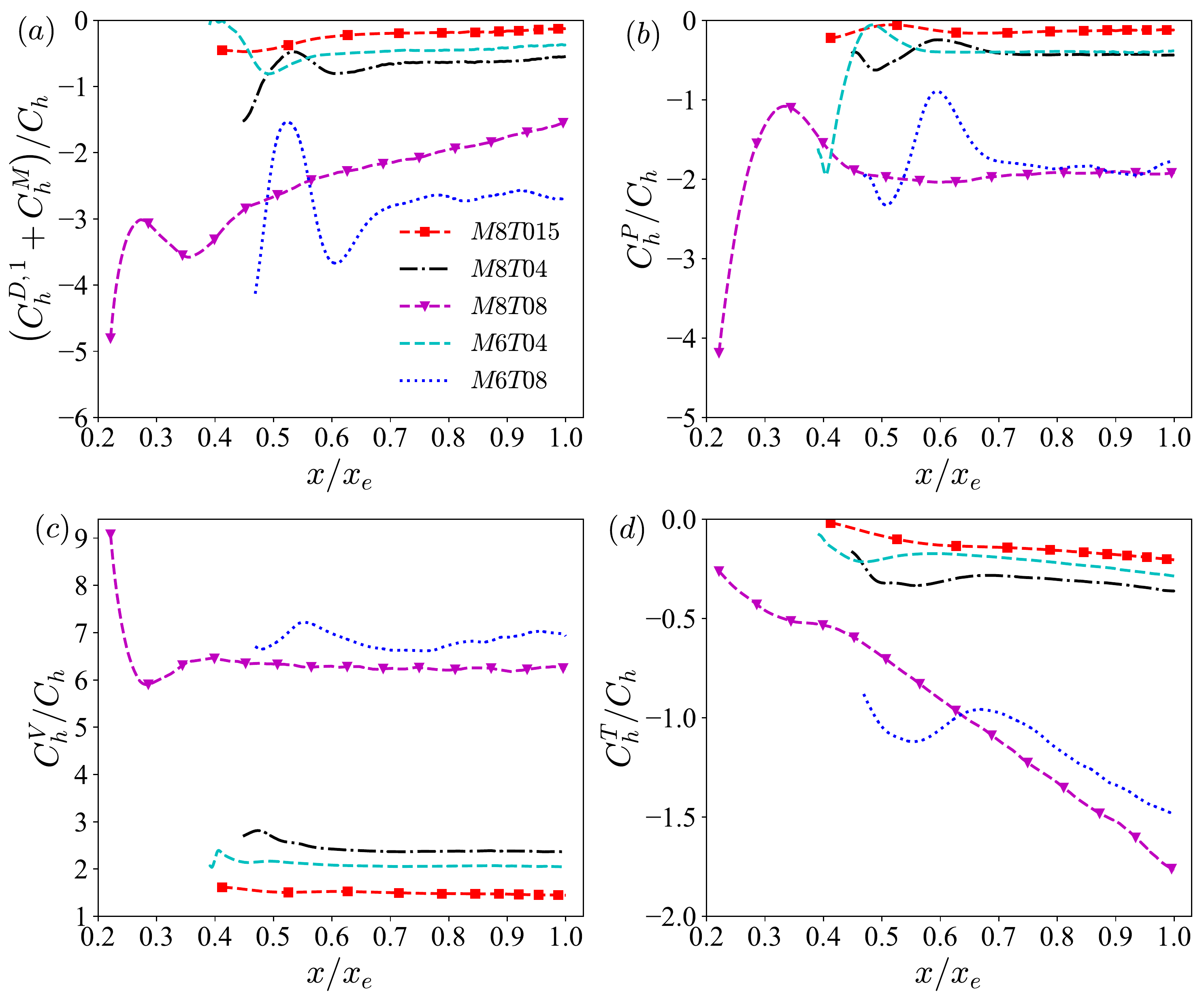}
	\caption{The streamwise evolutions of the relative contributions of (a) $C_{h}^{D,1}+C_{h}^{M}$, (b) $C_{h}^{P}$, (c) $C_{h}^{V}$ and (d) $C_{h}^{T}$ to the heat transfer coefficient $C_{h}$ in the hypersonic transitional and turbulent boundary layers.}
	\label{fig: d9}
\end{figure}

The streamwise evolutions of the relative contributions of $C_{h}^{D,1}+C_{h}^{M}$, $C_{h}^{P}$, $C_{h}^{V}$ and $C_{h}^{T}$ to the heat transfer coefficient $C_{h}$ in the hypersonic transitional and turbulent boundary layers with different Mach numbers and wall temperatures are shown in figure \ref{fig: d9}. On the contrary to the relatively weak correlation between the relative contributions to the skin-friction coefficient $C_{f}$ and the wall temperature, the relative contributions to the heat transfer coefficient $C_{h}$ is strongly influenced by the wall temperature. It is found that after the transition peak, the magnitudes of the relative contributions of $C_{h}^{P}$ and $C_{h}^{V}$ are almost constant, while that of $\left (C_{h}^{D,1}+C_{h}^{M}  \right )/C_{h}$ slightly decreases, and that of $C_{h}^{T}/C_{h}$ mildly increases along the streamwise direction. Moreover, the magnitudes of $\left (C_{h}^{D,1}+C_{h}^{M}  \right )/C_{h}$, $C_{h}^{P}/C_{h}$, $C_{h}^{V}/C_{h}$ and $C_{h}^{T}/C_{h}$ decrease significantly as the wall temperature decreases in the fully developed state of the hypersonic boundary layers. However, the larger magnitudes of these components cancel each other out and finally result in the smaller values of the heat transfer coefficient $C_{h}$ in weakly cooled cases.

\begin{figure}\centering
         \includegraphics[width=0.96\linewidth]{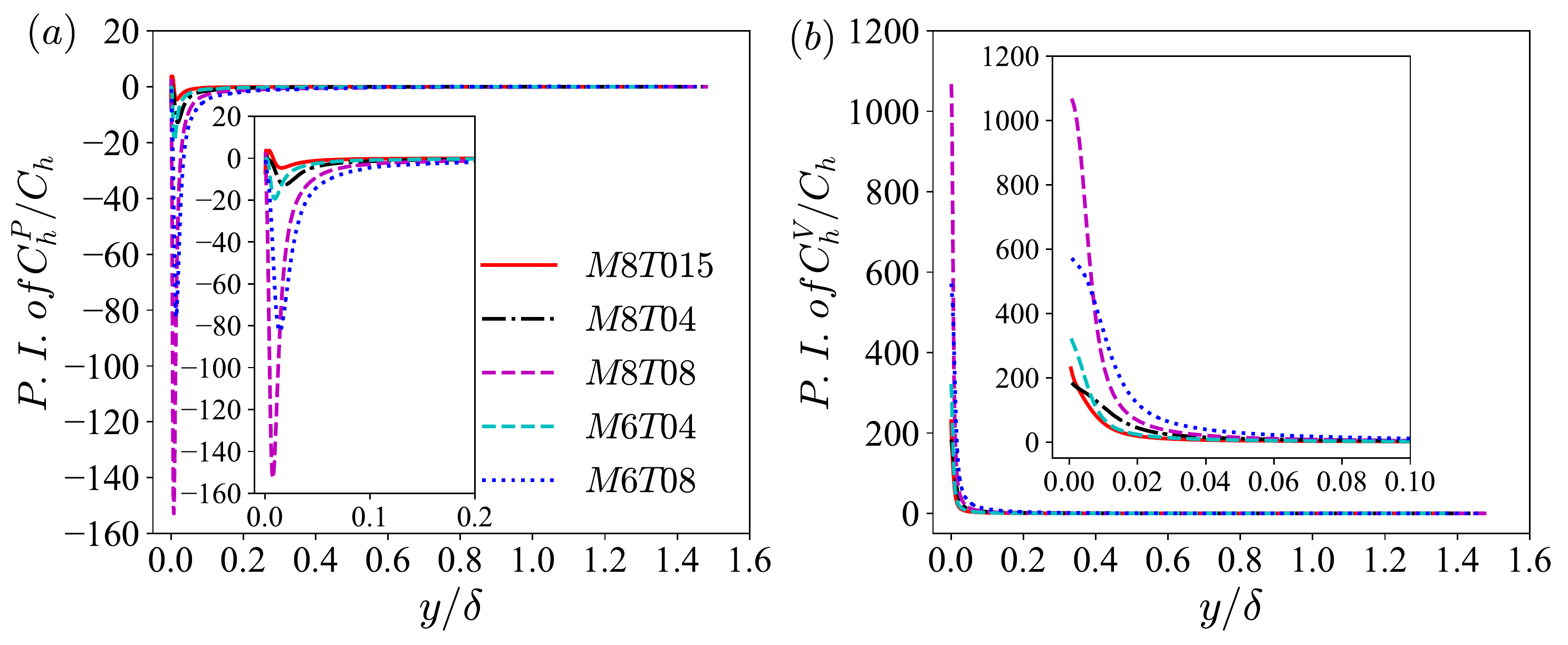}
	\caption{The pre-multiplied integrands of the relative contributions of (a) $C_{h}^{P}$ and (b) $C_{h}^{V}$ to the heat transfer coefficient $C_{h}$ in the hypersonic transitional and turbulent boundary layers at the streamwise position $x/x_{e}=0.95$. Here the abbreviation ``P. I.'' represents the pre-multiplied integrand. The pre-multiplied integrand of $C_{h}^{P}/C_{h}$ is expressed as $-\frac{C_{v}}{\left ( \overline{\rho} _{\delta }\overline{u}_{\delta }C_{p}\left ( T_{r}-T_{w} \right ) \right )C_{h}}\left ( y_{b}-y \right )\overline{\rho \left ( \gamma -1 \right )T\frac{\partial u_{j}}{\partial x_{j}}}$, and the pre-multiplied integrand of $C_{h}^{V}/C_{h}$ is defined as $\frac{C_{v}}{\left ( \overline{\rho} _{\delta }\overline{u}_{\delta }C_{p}\left ( T_{r}-T_{w} \right ) \right )C_{h}}\left ( y_{b}-y \right )\overline{\frac{\sigma _{ij}}{C_{v}Re}\frac{\partial u_{i}}{\partial x_{j}}}$.}
	\label{fig: d10}
\end{figure}

The pre-multiplied integrands of the relative contributions of $C_{h}^{P}$ and $C_{h}^{V}$ to the heat transfer coefficient $C_{h}$ in the hypersonic transitional and turbulent boundary layers at the streamwise position $x/x_{e}=0.95$ are shown in figure \ref{fig: d10}. It is found that the magnitudes of the pre-multiplied integrand of $C_{h}^{P}/C_{h}$ achieve their peak values in the buffer layer, indicating that the pressure-dilatation effect is dominant in the buffer layer. Furthermore, the peak values of the magnitudes of the pre-multiplied integrand of $C_{h}^{P}/C_{h}$ increase as the wall temperature increases, and the small positive peak values appear in the near-wall region only in the strongly cooled case (i.e. M8T015). The above observations can be explained as follows. According to the definition of the pressure-dilatation term in equation \ref{eqn: Cph}, if the velocity divergence $\partial u_{j}/\partial x_{j}$ is positive, the pre-multiplied integrand of $C_{h}^{P}/C_{h}$ is negative, and vice versa. It is noted that the mean flow in compressible turbulent boundary layers tends to be expansive as pointed out in \citet[]{Wang2012}; accordingly, the mean velocity divergence should be positive and the pre-multiplied integrand of $C_{h}^{P}/C_{h}$ should be negative across the turbulent boundary layer, as shown in figure \ref{fig: d9} (a). Furthermore, it was shown that the cold wall temperature can enhance the compression motions in the near-wall region in previous investigations \citep[]{Xu2021,Xu2021b}, thus the positive values of the mean velocity divergence should decrease as the wall temperature decreases across the turbulent boundary layer, and the mean velocity divergence can be negative near the wall due to strongly cooling of the wall (i.e. M8T015) \citep[]{Xu2021}. This further leads to the negative values of the pre-multiplied integrand of $C_{h}^{P}/C_{h}$ across the turbulent boundary layer, and the decrease of the negative values of the pre-multiplied integrand of $C_{h}^{P}/C_{h}$ as the wall temperature decreases. Moveover, the small positive peak values of the pre-multiplied integrand of $C_{h}^{P}/C_{h}$ in M8T015 are mainly due to the enhanced compression motions by the strongly cooling of the wall in the near-wall region.

It is shown in figure \ref{fig: d10} (b) that the pre-multiplied integrand of $C_{h}^{V}/C_{h}$ is dominant in the near-wall region, and sharply decreases away from the wall, which indicates that the effect of the viscous dissipation is dominant near the wall, and is consistent with the previous observation \citep[]{Xu2021b}. Moreover, the pre-multiplied integrand of $C_{h}^{V}/C_{h}$ decreases as the wall temperature decreases, which further results in the decrease of the $C_{h}^{V}/C_{h}$ with more cooled wall.

\begin{figure}\centering
         \includegraphics[width=0.96\linewidth]{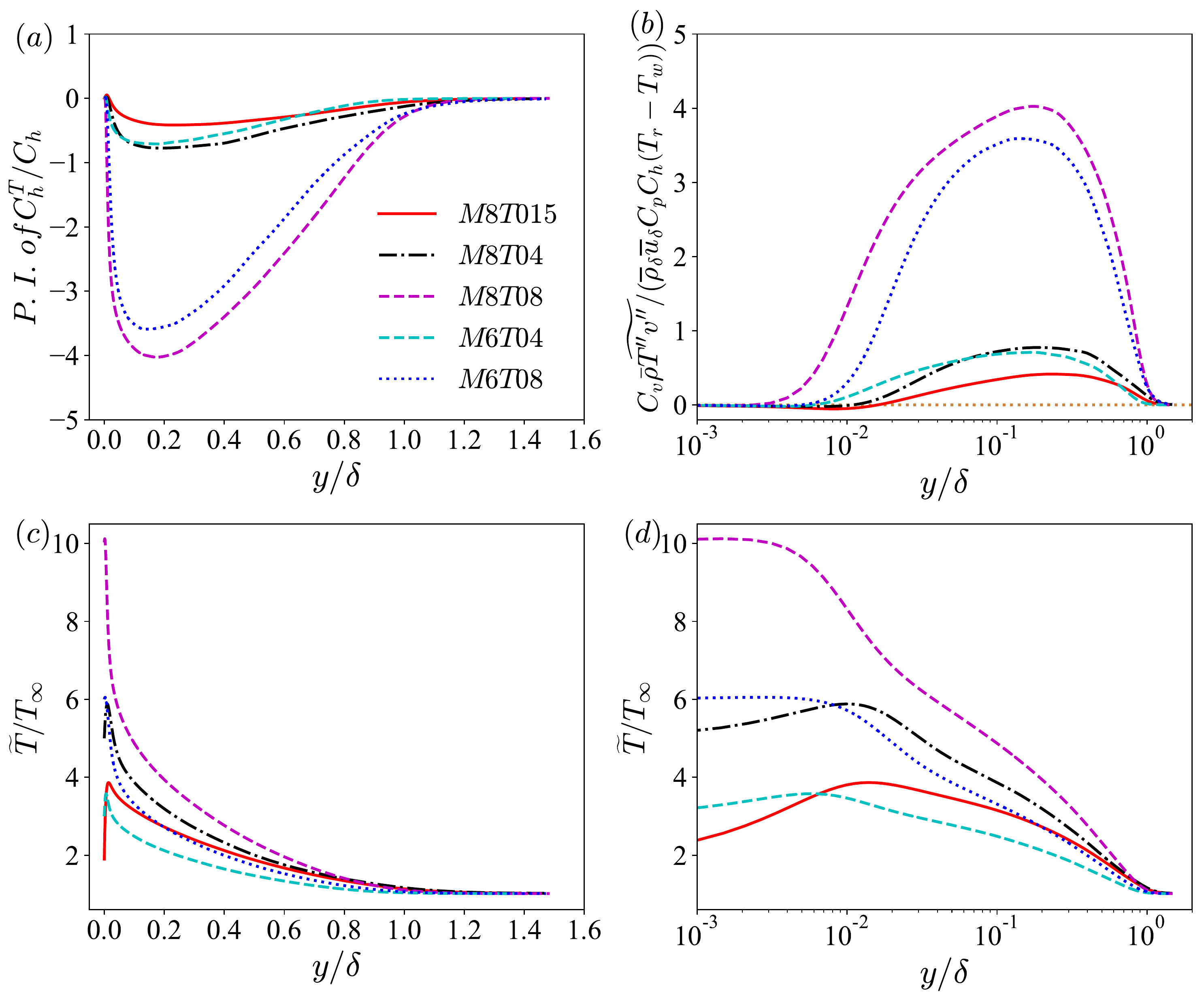}
	\caption{(a) The pre-multiplied integrand of the relative contribution of $C_{h}^{T}$ to the heat transfer coefficient $C_{h}$ in the hypersonic transitional and turbulent boundary layers at the streamwise position $x/x_{e}=0.95$. Here the abbreviation ``P. I.'' represents the pre-multiplied integrand. (b) The normalized Reynolds heat flux $C_{v}\bar{\rho }\widetilde{{T}^{\prime\prime}{v}^{\prime\prime}}/\left (\overline{\rho} _{\delta }\overline{u}_{\delta }C_{p}C_{h}\left ( T_{r}-T_{w} \right )  \right )$ along wall-normal direction at $x/x_{e}=0.95$. (c) and (d): The normalized mean temperature profiles $\widetilde{T}/T_{\infty }$ along wall-normal direction at $x/x_{e}=0.95$ with (c) linear-scaling $x$-axis and (d) log-scaling $x$-axis. The pre-multiplied integrand of $C_{h}^{T}/C_{h}$ is written as $-\frac{C_{v}\bar{\rho }\widetilde{{T}''{v}''}}{\left ( \overline{\rho} _{\delta }\overline{u}_{\delta }C_{p}\left ( T_{r}-T_{w} \right ) \right )C_{h}}$.}
	\label{fig: d11}
\end{figure}

The pre-multiplied integrand of the relative contribution of $C_{h}^{T}$ to the heat transfer coefficient $C_{h}$ and the normalized Reynolds heat flux $C_{v}\bar{\rho }\widetilde{{T}^{\prime\prime}{v}^{\prime\prime}}/\left (\overline{\rho} _{\delta }\overline{u}_{\delta }C_{p}C_{h}\left ( T_{r}-T_{w} \right )  \right )$ along wall-normal direction at the streamwise position  $x/x_{e}=0.95$ are shown in figure \ref{fig: d11} (a) and (b). It is shown that the normalized Reynolds heat flux is slightly negative in strongly cooled cases (including M8T015, M8T04 and M6T04) near the wall, and then switches to positive as the wall-normal distance increases; while for the weakly cooled cases (including M8T08 and M6T08), the normalized Reynolds heat flux is positive along the wall-normal direction. This can be explained as follows. According to the idea in Prandtl$^{'}$s mixing-length theory, the Reynolds heat flux $\bar{\rho }\widetilde{{T}^{\prime\prime}{v}^{\prime\prime}}$ is negatively correlated with the mean temperature gradient $\partial \widetilde{T}/\partial y$. The negative values of the Reynolds heat flux are correlated with the positive mean temperature wall-normal gradient near the wall in M8T015, M8T04 and M6T04, as shown in figure \ref{fig: d11} (c) and (d). As the wall-normal distance increases, the mean temperature wall-normal gradient becomes negative, and results in the positive Reynolds heat flux. The location of the peak values of the Reynolds heat flux is consistent with the location where the mean temperature wall-normal gradient is largest. Furthermore, the Reynolds heat flux decreases as the wall temperature decreases, which is mainly due to the reason that the mean temperature wall-normal gradient is smaller with colder wall temperature. It is noted that the pre-multiplied integrand of the relative contribution of $C_{h}^{T}$ to the heat transfer coefficient $C_{h}$ is opposite to the normalized Reynolds heat flux $C_{v}\bar{\rho }\widetilde{{T}^{\prime\prime}{v}^{\prime\prime}}/\left (\overline{\rho} _{\delta }\overline{u}_{\delta }C_{p}C_{h}\left ( T_{r}-T_{w} \right )  \right )$. Thus, on the contrary to the normalized Reynolds heat flux, the pre-multiplied integrand of the relative contribution of $C_{h}^{T}$ to the heat transfer coefficient $C_{h}$ is negative in most regions along wall-normal direction, which further results in the negative contribution to the heat transfer coefficient $C_{h}$. As the wall temperature decreases, the decrease of the Reynolds heat flux gives rise to the decrease of the relative contribution $C_{h}^{T}/C_{h}$.

\section{Summary and conclusion}\label{sec: n5}
In this paper, the decompositions of the skin-friction and heat transfer coefficients based on the two-fold repeated integration in hypersonic transitional and turbulent boundary layers with different Mach numbers and wall temperatures are analyzed to reveal the underlying mechanisms of the generations of the wall skin friction and heat transfer. To the best of our knowledge, it is the first time to investigate the decompositions of the skin-friction and heat transfer coefficients in transitional wall-bounded flows, and give the explanations to the overshoot of wall skin friction and heat transfer based on the integral analysis. It is found that the Reynolds analogy factor is small in the laminar state, and then increases due to the bypass transition process. An overshoot is also observed at the transition peak, which is similar to the behaviours of the skin-friction and heat transfer coefficients. It is also shown that the Reynolds analogy factor slightly increases as the wall temperature decreases, especially for the extremely cooled wall.

The values of the skin-friction coefficient in the hypersonic turbulent boundary layers increase as the wall temperature decreases. It is found that the mean velocity gradients, the Reynolds shear stress, the streamwise pressure gradient and the streamwise normal Reynolds stress gradient are important mechanisms during the bypass transition process. The overshoot of the skin-friction coefficient is mainly caused by the rapid change of the mean velocity profiles, rather than the Reynolds shear stress inferred by \citet[]{Franko2013}. In the hypersonic turbulent boundary layers, only the effects of the mean velocity gradients and the Reynolds shear stress contribute positively to the skin-friction coefficient, where the effect of the mean velocity gradients has the dominant contribution to $C_{f}$. Furthermore, the relative contributions of the mean velocity gradients and the Reynolds shear stress to the skin-friction coefficient have weak correlations with the wall temperature in hypersonic turbulent boundary layers, except for the strongly cooled wall condition. It is shown that the strongly cooled wall condition and high Mach number can enhance the effect of the Reynolds shear stress, and weaken the impact of the mean velocity gradients. Further investigation indicates that the smaller relative contribution of the mean velocity gradients in the strongly cooled case is mainly due to the large wall-normal gradient of the mean streamwise velocity $\partial \widetilde{u}/\partial y$ near the wall, while the larger relative contribution of the Reynolds shear stress is primarily caused by the enhancement of the normalized Reynolds shear stress by the strongly cooled wall.

Moreover, the heat transfer coefficient in the hypersonic turbulent boundary layers increases as the wall temperature decreases. The effect of the viscous dissipation gives the dominant positive contribution, while the mean temperature gradients, pressure dilatation and the Reynolds heat flux give the negative contributions to the heat transfer coefficient along the transitional and turbulent boundary layers. The overshoot of the heat transfer coefficient is primarily caused by the viscous dissipation, rather than the Reynolds heat flux inferred by \citet[]{Franko2013}. It is shown that the magnitudes of the relative contributions of the mean temperature gradients, pressure dilatation, viscous dissipation and the Reynolds heat flux increase as the wall temperature increases in the hypersonic turbulent boundary layers. It was shown that the cold wall temperature can enhance the compression
motions in the near-wall region in previous investigations \citep[]{Xu2021,Xu2021b}, which further leads to the decrease of the magnitude of the relative contribution of the pressure dilatation with colder wall temperature. It is also found that the effect of the viscous dissipation is dominant near the wall, and sharply decreases away from the wall. Furthermore, the decrease of the relative contribution of the Reynolds heat flux is mainly due to the smaller mean temperature wall-normal gradient with colder wall temperature.

In conclusion, the underlying mechanisms of the generations of the wall skin friction and heat transfer are revealed in the hypersonic transitional and turbulent boundary layers. The reasons of the overshoot of the skin-friction and heat transfer coefficients are given based on the integral analysis, which could be helpful for the skin friction reduction and thermal protection in hypersonic aircrafts.

\acknowledgments{
\noindent\textbf{Funding.} This work was supported by the NSFC Basic Science Center Program (Grant No. 11988102), by National Natural Science Foundation of China (NSFC Grants No. 91952104, 92052301, 12172161 and 91752201), by the Technology and Innovation Commission of Shenzhen Municipality (Grant Nos. KQTD20180411143441009 and JCYJ20170412151759222), and by Department of Science and Technology of Guangdong Province (Grant No. 2019B21203001). This work was also supported by Center for Computational Science and Engineering of Southern University of Science and Technology.

\noindent\textbf{Declaration of Interests.} The authors report no conflict of interest.}

\bibliographystyle{jfm}
\bibliography{jfm-instructions}

\end{document}